\documentclass[12pt,preprint]{emulateapj}

\def\ro{{\it ROSAT\/}}
\def\asca{{\it ASCA\/}}
\def\xmm{{\it XMM-Newton\/}}
\def\cha{{\it Chandra\/}}
\def\cgps{{\it CGPS\/}}

\shorttitle{\xmm, \cha, \& \cgps\ observations of G85.4+0.7 and G85.9$-$0.6}
\shortauthors{Jackson, Safi-Harb, Kothes, \& Foster}

\begin{document}

\title{\xmm, \cha, and \cgps\ observations of the Supernova Remnants G85.4+0.7 and G85.9$-$0.6}

\author{M. S. Jackson\altaffilmark{1}, S. Safi-Harb\altaffilmark{1,2,3}, 
R.
Kothes\altaffilmark{4, 5}, T. Foster\altaffilmark{6}}
\altaffiltext{1}{Department of Physics and Astronomy, University of
Manitoba, Winnipeg, MB, R3T 2N2, Canada.}
\altaffiltext{2}{Canada Research Chair}
\altaffiltext{3}{Department of Physics, George Washington University,
Washington, DC 20052, U.S.A.}
\altaffiltext{4}{National Research Council of Canada, Herzberg Institute
of Astrophysics, Dominion Radio Astrophysical Observatory, P.O. Box 248,
Penticton, BC V2A 6J9, Canada.}
\altaffiltext{5}{Department of Physics and Astronomy, University of
Calgary, 2500 University Drive NW, Calgary, AB T2N 1N4, Canada.}
\altaffiltext{6}{Department of Physics \& Astronomy, Brandon University,
270-18th Street, Brandon, MB, R7A 6A9 Canada.}

\begin{abstract}

﻿We present an \xmm\ detection of two low radio surface brightness
SNRs, G85.4+0.7 and G85.9$-$0.6, discovered with the Canadian Galactic
Plane Survey (CGPS). High-resolution \xmm\ images revealing the morphology
of the diffuse emission, as well as discrete point sources,
are presented and correlated with radio and \cha\ images. The new data also permit a
spectroscopic analysis of the diffuse emission regions, and a spectroscopic
and timing analysis of the point sources. Distances have
been determined from \ion{H}{1} and CO
data to be $3.5\pm 1.0$ kpc for SNR G85.4+0.7 and $4.8\pm 1.6$ kpc for SNR G85.9$-$0.6.

The SNR
G85.4+0.7 is found to have a temperature of $\sim12-13$ MK and a 0.5--2.5~keV
luminosity of $\sim1-4\times 10^{33}D^2_{3.5}$ erg/s (where $D_{3.5}$ is
the distance in units of 3.5 kpc), with an electron
density $n_e$ of $\sim0.07-0.16(fD_{3.5})^{-1/2}$ cm$^{-3}$ (where $f$ is
the volume filling factor), and a
shock age of $\sim9-49(fD_{3.5})^{1/2}$ kyr.

The SNR
G85.9$-$0.6 is found to have a temperature of $\sim15-19$ MK and a 0.5--2.5~keV
luminosity of $\sim1-4\times 10^{34}D^2_{4.8}$ erg/s (where $D_{4.8}$ is
the distance in units of 4.8 kpc), with an electron
density $n_e$ of $\sim0.04-0.10(fD_{4.8})^{-1/2}$ cm$^{-3}$ and a
shock age of $\sim 12-42(fD_{4.8})^{1/2}$ kyr.

Based on the data presented here, none of the point
 sources appears to be the neutron star associated with either SNR.

\end{abstract}


\section{Introduction}

In 2001, two new supernova remnants (SNRs) with low radio surface
brightness, G85.4+0.7 and G85.9$-$0.6,  were
discovered in Canadian Galactic Plane Survey (CGPS)\citep{tay03} data and confirmed
in X-rays with \ro\ data \citep{kot01}. Both show distinct shells in
the radio band with an extended region of X-ray emission in the
centre. The radio surface brightness of G85.4+0.7 at 1 GHz is
$\Sigma_{1 {\rm GHz}}\le 1\times 10^{-22}$ Watt ${\rm m}^{-2}$ ${\rm Hz}^{-1}$
${\rm sr}^{-1}$ and the radio data also indicate that the SNR has a non-thermal shell
with angular diameter $\approx0.4^{\circ}$ which is surrounded by a
thermal shell with an angular diameter of $\approx0.6^{\circ}$ and is
located within an H I bubble. The bubble also contains two B stars which may have been part of the
same association as the SNR's progenitor star. G85.9$-$0.6 has a radio surface brightness
$\Sigma_{1 {\rm GHz}}\le 2\times 10^{-22}$ Watt ${\rm m}^{-2}$ ${\rm Hz}^{-1}$
${\rm sr}^{-1}$ and it has no discernible H I features, indicating
that it is expanding into a low density medium, perhaps between the
local and Perseus spiral arms. The most likely event which would
produce an SNR in such a region would be a Type Ia supernova.

X-ray observations are important to the study of SNRs, particularly those with
low surface brightness, because they provide information about the
morphology and emission processes of these objects, which are
indicators of both the nature of the supernova explosion which formed them
and the properties of the progenitor star. SNRs
with low surface brightness are expected to be formed after the core
collapse of a massive star in a Type Ib/c or Type II explosion,
because the stellar wind would have blown away much of the
interstellar medium (ISM) surrounding it, leaving a low ambient
density into which the shock from the supernova expands. A Type Ia supernova can also
result in a low surface brightness SNR if the surrounding ISM has a low density, such as 
it would if it were located between two spiral arms of the galaxy. 
Thermal X-ray emission from a SNR arises as the blast wave of
the explosion travels through and shocks the ISM, and as
a reverse shock travels back into and shocks the ejecta.
The X-ray
spectrum of the SNR gives information about the temperature, the
density, and the luminosity of the shocked material, while imaging
data provides information about the size and morphology of the region.

The low
angular and spectral resolutions as well as the small number of counts
in the \ro\ data did not allow spectroscopy nor detailed imaging to be 
done, so \xmm\
observations, described in \S\ref{obs}, have been made in order to confirm the
detection of the SNRs, and to perform imaging, spectroscopic and
timing studies. Detailed X-ray imaging, described in \S\ref{imaging},
is used to map the diffuse emission and compare it to the
location and size of the radio shells, and \cha\ data have been used to
search for compact objects not resolved by XMM. Spectral parameters obtained in
\S\ref{spec} lead to an
estimate of such quantities as temperature, density, and luminosity of the
SNRs. In \S\ref{ps} the point sources are catalogued and an attempt at
identification is made by matching their positions to objects in other
catalogues. Timing analysis is performed to identify any pulsar
candidates. The distances to the SNRs are derived from \ion{H}{1} and
CO data in
\S\ref{dist}. The results are discussed in \S\ref{disc}.

Preliminary results were presented in \cite{jac06} and \cite{saf06}.

\section{Observations}\label{obs}

\subsection{\xmm\ Observations}
SNR G85.4+0.7 was observed with \xmm\ on 2005 May 31 for 11.5 ks (Obs
ID: 0307130101, PI: S. Safi-Harb) and again on
October 27 for 15.2 ks (Obs ID 0307130301, PI: S. Safi-Harb), because
of proton flares in the first 
observation which made a large fraction of the data unusable. A 29 ks
\xmm\ observation was made of SNR G85.9$-$0.6 on 2005 November 24 (Obs
ID 0307130201, PI: S. Safi-Harb). 

The PN \citep{str01} and MOS \citep{tur01} data were reduced with the latest version of SAS (6.5.0) and
events during proton flares were filtered out for producing images and
spectra by using the SAS routines {\it evselect} to create rate files
and {\it tabgtigen} to generate good time intervals (GTIs) for
filtering the event files with the {\it evselect} routine.  This rendered the
first observation of G85.4+0.7 useful only
for imaging, and slightly reduced the integration time, from 29 ks to 26 ks for the
G85.9$-$0.6 observation. The second observation of G85.4+0.7 shows no
evidence for proton flaring and the removal of proton flares did not 
significantly reduce the integration time. Images and spectra were created from the
event files using {\it evselect}. The total effective exposure time
for G85.4+0.7 was 14~ks for PN and 16~ks for the MOS instruments and
for G85.9$-$0.6 it was 23~ks for PN and 26~ks for MOS.

To facilitate source detection, exposure maps were made using {\it
  eexpmap} and the images were combined using {\it emosaic}. The
  spectra were binned using a minimum of 25-50 counts per bin for the
MOS instruments and 50 or 100 counts per bin for the PN for the point source 
and diffuse spectra respectively, using {\it grppha}. This latter
grouping was necessary because the background subtraction added
excessive noise to the spectra, particularly at high energies, where
the spectra are background-dominated. The background for both SNRs
was calculated using various methods described in \S\ref{bg}.

\subsection{\cha\ Observations}
A 14.5 ks observation of SNR G85.4+0.7 was made on 2003 January 26
(Obs ID: 3898, PI: S. Safi-Harb) by the 
Advanced CCD Imaging Spectrometer (ACIS-S; G. Garmire\footnote{See 
http://cxc.harvard.edu/proposer/POG.}).
SNR G85.9$-$0.6 was observed with ACIS-S for 20.2 ks on 2003 October
3 (Obs ID: 3899, PI: S. Safi-Harb). Both observations 
were made at a focal plane temperature of $-120^{\circ}$C.

The analysis of the \cha\ data was done using Chandra Interactive
Analysis of Observations (CIAO version 3.3\footnote{http://cxc.harvard.edu/ciao}). For both \cha\
observations, the data were corrected for charge transfer inefficiency 
(CTI) with tools provided by the ACIS team of Pennsylvania \citep{tow00}. Events with \asca\ grades 
(0, 2, 3, 4, 6) were retained, and periods of high background rates were removed. Data 
from hot pixels were eliminated.
The effective exposure time for the observation of SNR G85.4+0.7 was 14.3 ks and for 
SNR G85.9$-$0.6 it was 19.9 ks. The backgrounds for the point source spectra were 
extracted from annuli surrounding the sources. The spectra were grouped with a mimimum of 30 counts in each bin.
Data from the S2 and S3 chips were used. Data from the other chips were not used because they do not overlap the XMM field of view.

\section{Imaging}\label{imaging}

The previous X-ray images from the \ro\ All-Sky Survey did not resolve
the SNRs and the point sources in the field, so the \xmm\ PN and
MOS data can be used to determine the nature of the sources. The
results are described below.

\subsection{G85.4+0.7}
\begin{figure}
\scalebox{1.0}{\rotatebox{0}{\plotone{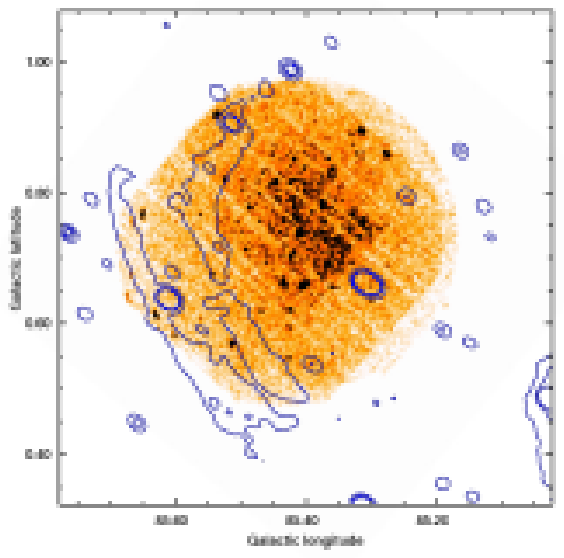}}}
\caption{0.5--2.5 keV \xmm\ X-ray mosaic image of G85.4+0.7 with the
  highest few radio contours overlaid in blue. The X-ray image is Gaussian smoothed with a radius
  of 3 pixels. 
\label{85.4image}}
\end{figure}

The 0.5--2.5 keV \xmm\ PN and MOS mosaic image of G85.4+0.7 with radio contours overlaid is 
shown in Figure~\ref{85.4image}. The image includes only energies
between 0.5 and 2.5 keV in order to match the ROSAT bandpass and
because there is negligible diffuse emission above 2.5 keV. The X-ray
image is smoothed with a
Gaussian filter with a 3 pixel radius. The X-ray emission is in the same
location and is a similar angular size as indicated by the \ro\ data, shown 
in Figure 6 of \cite{kot01}, with an approximate angular radius of
6.7$^{\prime}$. It is much better resolved, however, and
the regions of diffuse emission and the point sources are clearly visible.

\begin{figure}
\center{\scalebox{0.45}{\rotatebox{0}{\plotone{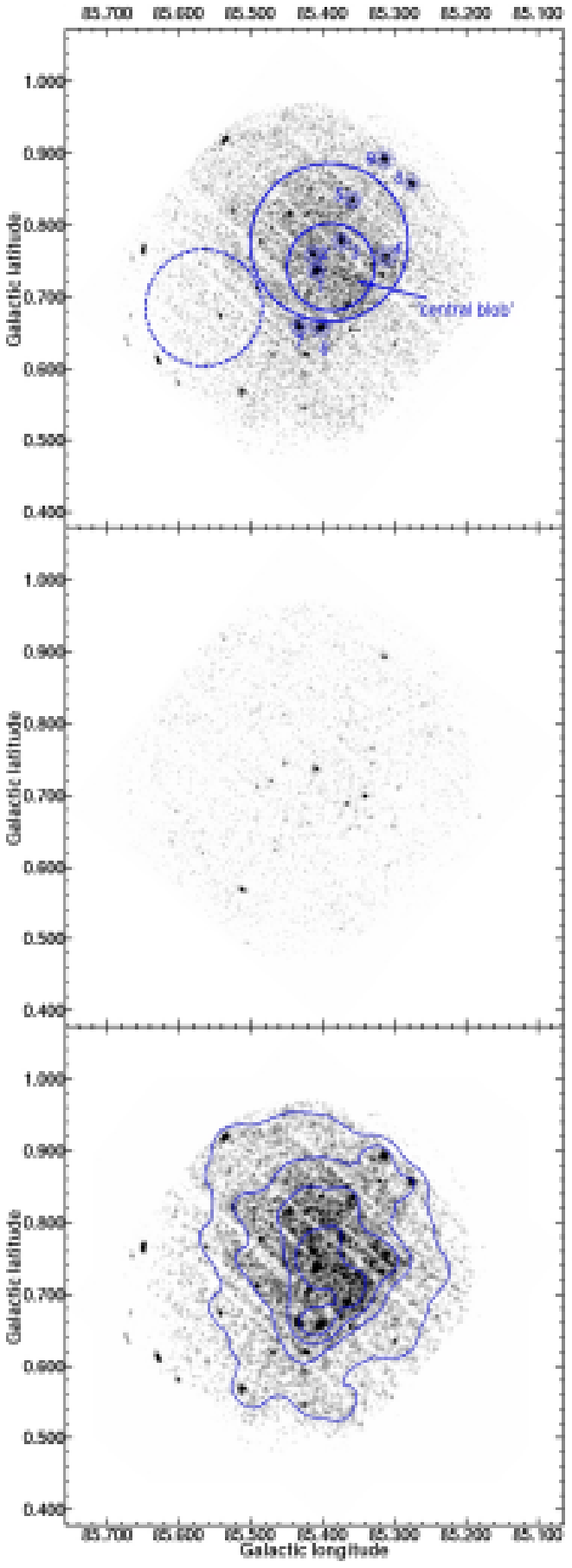}}}}
\caption{PN and MOS mosaic images of G85.4+0.7. Top panel: 0.5--2.5
  keV image, showing point
  sources (1-9) and the diffuse emission extraction region, as well as the region used for the central blob (see \S\ref{disc85.4}).
 The
  background region for the diffuse spectra is also shown
  as a dashed circle. The image is smoothed as in
  Figure~\ref{85.4image}. Middle panel: 2.5--10.0 keV image, similarly smoothed
  and with the same contrast as in the top panel. Bottom panel: 0.5--2.5
  keV image, Gaussian smoothed with a radius of 3 pixels and
  logarithmically scaled and adjusted in contrast
  to emphasize the diffuse emission and overlaid with
  contours of the 0.5--2.5 keV emission smoothed similarly to the \ro\
  contours in Figure~6 of
  \cite{kot01}.\label{85.4reg}} 
\end{figure}

The top two panels of Figure~\ref{85.4reg} show the 0.5--2.5~keV and 
2.5--10~keV \xmm\ images of G85.4-0.7, smoothed in the same way as 
Figure~\ref{85.4image}. The absence of diffuse emission in the hard X-ray 
band is clear from the middle panel. The lower panel shows the 
0.5--2.5~keV image with sources subtracted, smoothed further and scaled to emphasize the diffuse 
emission, and smoothed contours are overlaid to show the morphology of the 
remnant. The 
contours indicate a centrally filled morphology with a slightly elongated 
shape. The circles in the top panel indicate spectral extraction regions described in \S\ref{spec85.4} and \S\ref{disc85.4}.

There are nine point sources in the image from which spectra can be
extracted, and the analysis is described in \S\ref{ps}. The extraction
regions for the spectra, including the
background for the diffuse emission, are indicated and labeled in
Figure~\ref{85.4reg}, and the spectral analysis of the diffuse
emission is described in \S\ref{spec85.4}.

The \cha\ image of SNR G85.4+0.7 is shown in Figure~\ref{cha85.4}. The extraction 
regions used for the source and background regions are shown, and the 
applicable regions from the \xmm\ observation are overlaid. The
{\it wavdetect} tool in CIAO is used for source detection. Six of the point sources 
identified with \xmm\ (1, 2, 3, 4, 6 and 7) are clearly visible on the \cha\ image. Twelve 
additional sources appear on the \cha\ image (labeled 10$-$21) but the spectra 
extracted from these point sources do not contain enough counts to enable meaningful 
spectral analysis, so while they are potential neutron star
candidates and sources 10, 11, 12, 13, 15, and 16 are situated well within the radio shell of the SNR,
further investigation of these sources is not possible at
this time.

\subsection{G85.9$-$0.6}
The \xmm\ PN and MOS mosaic image of G85.9$-$0.6 with radio contours overlaid is shown in
Figure~\ref{85.9image}. The image is smoothed in the same way as the
G85.4+0.7 image, with a Gaussian filter with a three-pixel radius. As with
G85.4+0.7, the angular radius of approximately 5.8$^{\prime}$ and 
location of the X-ray emission shown in this figure matches that of
Figure 6 of \cite{kot01} and many features are resolved. 0.5--2.5 keV
and 2.5--10.0 keV X-ray images are shown in the top two panels of
Figure~\ref{85.9reg}. It is clear from the hard X-ray image that the
SNR does not appear above 2.5 keV, and the X-ray spectrum is 
background dominated above 2.5 keV, as will be explored in 
\S\ref{spec85.9}. The bottom panel of Figure~\ref{85.9reg} shows the same 
data as in the top panel with point source emission subtracted, scaled to emphasize
the diffuse emission. The contours in the bottom panel are smoothed
to show the overall morphology of the remnant. The diffuse X-ray emitting region seems to be round
in shape, indicating a centrally filled X-ray morphology.
The circles in the top panel indicate spectral extraction regions described in \S\ref{spec85.9} and \S\ref{disc85.9}.

Spectra were extracted from 8 point sources and the diffuse region, as
indicated in the top panel of Figure \ref{85.9reg}. The
analysis of the spectrum from the diffuse region is described in
\S\ref{spec85.9}, and
\S\ref{ps} describes the analysis of the point sources.

Figure~\ref{cha85.9} shows the \cha\ image of SNR G85.9$-$0.6. The locations 
of the extraction circles for sources 1 and 8 in the \xmm\ image are shown, in addition to the additional point sources found on the \cha\ image (labeled 9$-$27). The \cha\ 
image does not show any point sources which have sufficient counts to do independent spectroscopic analysis, and only one of the sources identified from the \xmm\ image (source 8) lies 
within the field of view of the \cha\ image, with which a \cha\ source has been identified.

\section{Spectroscopy}\label{spec}

The spectra from the regions of diffuse emission were extracted from
areas indicated in Figures~\ref{85.4reg} and \ref{85.9reg} for both PN and
MOS. The
diffuse emission regions and point sources are indicated and labeled
in the Figures. 

The spectra of the diffuse emission for both SNRs were fit to various
models: a simple bresstrahlung model with gaussian lines,  VMEKAL, VNEI, and VPSHOCK, each modified by interstellar absorption, in this case the Wisconsin absorption model (wabs in XSPEC, \cite{morr83}). VMEKAL is a collisional ionization equilibrium model based on the 
model calculations of \cite{mew85} of the emission spectrum 
from hot diffuse gas with Fe calculations by \cite{lie90}, with variable abundances. VPSHOCK is a 
non-equilibrium ionization (NEI) model with variable abundances which 
models a plane parallel shock in a plasma with constant electron 
temperature T and a range of ionization timescales with an upper limit
$\tau=n_et$, where $n_e$ is the postshock electron density and $t$ is the 
shock age \citep{bor01}. VNEI is a NEI model similar to VPSHOCK, except with a single ionization timescale.

\subsection{Background Estimation}\label{bg}
Spectra of diffuse faint regions require particular attention to be
paid to the choice of background region. Instrumental emission lines
can dominate the spectrum if the background region is poorly chosen,
leading to errors in spectral fits and a discrepancy between the PN and
MOS spectra. The intuitive background region would be a relatively
source free one at approximately the same galactic latitude as the 
source to minimize contamination by the Galactic ridge. However, because 
the effect of
instrumental emission lines is not uniform with position on the CCDs,
particularly on the PN instrument
\citep{lum02}, and exceeds the variation with galactic latitude for
these observations, the background regions were chosen as closely as possible to the point
opposite the source region through the centre of the image, taking
care that there is no overlap between the source and background
regions and omitting regions around point sources. The background regions for both SNRs are necessarily smaller than the source regions, but very little change in the spectral fits resulted from choosing various background regions in the same general area of the image.

An attempt has been made to estimate the MOS background spectra using the
\xmm\ Extended Source Analysis Software (XMM-ESAS) \citep{snow06}. As
yet, background estimation for the PN instrument is not included in
this package. Data from filter wheel closed observations and from the
unexposed corners of the MOS CCDs are used to generate background
spectra. This serves as a test of the backgrounds generated from the
observation itself, described above. To fit spectra using the XMM-ESAS
backgrounds, it 
is necessary to include some additional spectral components, as
recommended in \cite{snow06}. Unabsorbed Gaussian lines of zero width
at energies of 1.49 and 1.75 keV are fit to the MOS spectra in
addition to the model describing the emission from the SNR. The 1.75
keV line was also added to the fit to the MOS spectra using the
observation background. It was not necessary to add the 1.49 keV line to the fit of the G85.4+0.7 MOS spectra using observation backgrounds 
because it was adequately subtracted, but the 1.49 keV line was added to the G85.9$-$0.6 MOS spectra using observation backgrounds. It was also
necessary to include a low energy broken power law component to the
G85.9$-$0.6 fit of the ESAS-subtracted spectrum, as instructed in the ESAS documentation, to correct for
instrumental effects, because the statistics allowed residuals at low energies
to be clearly seen in the ESAS background subtracted
spectra. The broken power law was not added to the spectrum of
G85.4+0.7 because the spectra were noisy and did not require an additional model
component. Results of spectral fits using both background estimation
methods for each SNR are given in Table~\ref{dfit}. The parameters in
the left hand column of Table~\ref{dfit} for each of the SNRs are
from the simultaneous fit to the MOS spectra using the ESAS background subtraction
method and the PN spectrum with the background spectrum extracted from
the observation itself, and the right hand column contains the
parameters from the simultaneous fit to the MOS and PN spectra from
both of which has been subtracted the observation background.

As a third background subtraction method, we have attempted to use blank sky event files that are available for the PN and MOS instruments and generated to enable the 
production of background spectra for extended sources. These event files comprise 
a superposition of pointed observations from which sources 
have been removed. Background files produced from blank sky event files therefore 
simulate the detector response in an actual observation, including any instrumental emission lines. Events are first 
selected from the original event files based on sky position using the {\it SelectRADec} script.
The {\it skycast}
script \citep{rea03} is then used to cast the new event file onto the sky coordinates 
for the particular observation. The background spectra can be extracted from these files from the same
region as for the source spectra. The BACKSCAL keyword is adjusted in the background file
to scale it to the source file. Unfortunately, the resulting background subtracted spectra 
are oversubtracted for the PN instrument, and contain negative 
counts below 0.6 keV and above 1.5 keV. Thus the blank sky background spectra cannot be used for this analysis. Possibly the 
oversubtraction results from the fact that there are no observations within $30^{\circ}$ 
of the SNRs G85.4+0.7 or G85.9$-$0.6 contained in the blank sky datasets, and therefore 
systematic errors arise from the different level of background emission. The background 
subtracted PN spectra of G85.4+0.6 using a background from the observation itself and from the blank sky
file is shown in Figure~\ref{PNback}. The spectra from which the blank
sky background was subtracted were not used and thus the parameters do
not appear in Table~\ref{dfit}.

\subsection{G85.4+0.7}\label{spec85.4}

The PN and MOS spectra between 0.5 and 2.5 keV are simultaneously fit in 
order to determine
the best parameters while reducing instrument-specific and
systematic effects. The upper limit of 2.5 keV is used for the spectral fits because the spectra are background dominated above this energy as shown in Figure~\ref{PNback}. 

Fitting to the VMEKAL model results in a reduced $\chi^2$ value of 1.76 for the ESAS background or 1.63 for 
the observation background (this refers to the MOS spectra
only; the background spectra extracted from the observation was used
for every PN spectrum), and a $kT$ value of $0.65\pm0.03$ or $0.65\pm0.01$ keV (all errors are $2\sigma$) 
for the ESAS and observation 
background respectively.

The spectra are best fit with an absorbed VPSHOCK
with with $kT$ of 1.1 $^{+0.8}_{-0.3}$ keV  for the ESAS background or $1.0^{+0.4}_{-0.2}$ keV for the
backgrounds extracted from the observation itself, and ionization timescale of
$(6.4^{+5.8}_{-2.8})\times 10^{10}$ cm$^{-3}$ s for the ESAS background or $(8.0^{+8.4}_{-3.4})\times
10^{10}$ cm$^{-3}$ s for the background extracted from the observation. The 
spectrum for the diffuse region is shown with the ESAS background in
Figure~\ref{85.4esasspec} and with the observation background in Figure~\ref{85.4spec}.
The parameters for the simultaneous PN and MOS fits, with 2$\sigma$ uncertainties, are given in
Table~\ref{dfit}, along with 
derived quantities such as luminosity and age. The fitted parameters
agree well within uncertainty for the two background estimation
methods. The abundances are for the most part consistent with solar,
with exceptions given in Table~\ref{dfit}. An analysis of the
spectral fit parameters has shown G85.4+0.7 to have a 0.5--2.5 keV
luminosity of $(3.1^{+1.3}_{-1.5})\times 10^{33}$ erg s$^{-1}$ or $(2.0^{+1.6}_{-1.0})\times 10^{33}$
erg s$^{-1}$ for the ESAS or observation backgrounds respectively, in
both cases based on the estimated distance of $3.5\pm 1.0$ kpc for
G85.4+0.7, determined in \S\ref{dist}. The
normalizations of the instrumental lines in the MOS spectra from which the
ESAS background has been subtracted are (1.7$\pm$0.2)$\times 10^{-5}$ and
(5.5$^{+0.9}_{-1.0}$)$\times 10^{-5}$ photons cm$^{-2}$s$^{-1}$ for the lines at 1.49 and 1.75 keV respectively. The normalization for the 
line at 1.75 keV for the observation background is ($3.7^{+1.0}_{-1.1}$)$\times10^{-5}$ photons cm$^{-2}$s$^{-1}$.

To verify the fits to elemental abundances, the VNEI model 
was used in place of VPSHOCK. All 
elemental abundances were found to be consistent within error to be solar 
or subsolar, except O and Fe. O was clearly enhanced above a solar value, 
giving strength to the VPSHOCK result. The elemental abundances using the VMEKAL model were also qualitatively similar to the VPSHOCK result.

\subsection{G85.9$-$0.6}\label{spec85.9}
The approach used for fitting the diffuse spectra of G85.9$-$0.6 is similar to that
used for G85.4+0.7. Again, the upper limit of 2.5 keV is used for the spectral fits because the spectra are background dominated above this energy. The PN diffuse spectrum was fitted simultaneously
with the MOS spectra from which either the ESAS or observation
background was subtracted. 

Fitting to the VMEKAL model results in a reduced $\chi^2$  of 2.21 for the ESAS background and 2.48 
for the observation background, a $kT$ value of $0.70\pm0.01$ or $0.68\pm 0.01$ keV.

The spectra of the diffuse emission from the remnants are best fit with
an absorbed VPSHOCK model with $kT$ of 
$1.3^{+0.5}_{-0.2}$ keV
for the ESAS background or $1.6^{+0.5}_{-0.3}$ keV for the
background extracted from the observation and ionization timescale $(6.8^{+1.9}_{-0.9}) \times
10^{10}$ cm$^{-3}$ s for the ESAS background or $(5.1^{+1.5}_{-1.0})\times 10^{10}$ cm$^{-3}$ s
for the background extracted from the observation. 
The fitted parameters with 2$\sigma$ uncertainties
are given in Table~\ref{dfit} and the diffuse spectrum with the ESAS
background is shown in
Figure~\ref{85.9esasspec} and with the observation background in
Figure~\ref{85.9spec}.
The abundances are for the most part consistent
with solar, and the exceptions are shown in Table~\ref{dfit}. The
luminosity of G85.9$-$0.6 is $(2.4^{+1.2}_{-1.1})
\times 10^{34}$ erg s$^{-1}$ for the ESAS background or $(2.7^{+1.2}_{-1.3}) \times 10^{34}$ erg s$^{-1}$
for the observation background, in both cases
based on the estimated distance of $4.8\pm1.6$ kpc for G85.9$-$0.6, determined in \S\ref{dist}. The
normalizations for the 1.49 and 1.75 keV lines in the ESAS subtracted
MOS spectra are (1.22$\pm0.08$)$\times 10^{-4}$ and
(3.7$^{+0.5}_{-0.6}$)$\times 10^{-5}$ photons cm$^{-2}$s$^{-1}$ respectively.
The broken power law which was added to the fit of the MOS ESAS-background
subtracted spectrum had $\Gamma_1=1.2^{+0.4}_{-0.5}$, $\Gamma_2=3.3^{+0.4}_{-0.3}$, $E_{\rm
  break}=0.91\pm0.05$ keV, and a normalization of
$(8.1^{+1.4}_{-1.1})\times 10^{-4}$
photons$/$keV$\cdotp$cm$^2\cdotp$ s at 1 keV. The normalization of the
1.75 keV line for the observation background is ($3.2\pm1.1$)$\times10^{-5}$ photons cm$^{-2}$s$^{-1}$.

When the VNEI model was used in place of VPSHOCK to check the elemental abundances, as was done with G85.4+0.7, all 
elemental abundances were found to be consistent within error to be solar 
or subsolar, except O and Fe, both of which are clearly above solar abundance
when the error bar is taken into account, in agreement with the VPSHOCK result. Using the VMEKAL model, the 
elemental abundances are again qualitatively similar to those obtained from the VPSHOCK model, except Mg is above solar.

\section{Point Source Analysis}\label{ps}
\begin{figure}
\center{\scalebox{1.}{\rotatebox{0}{\plotone{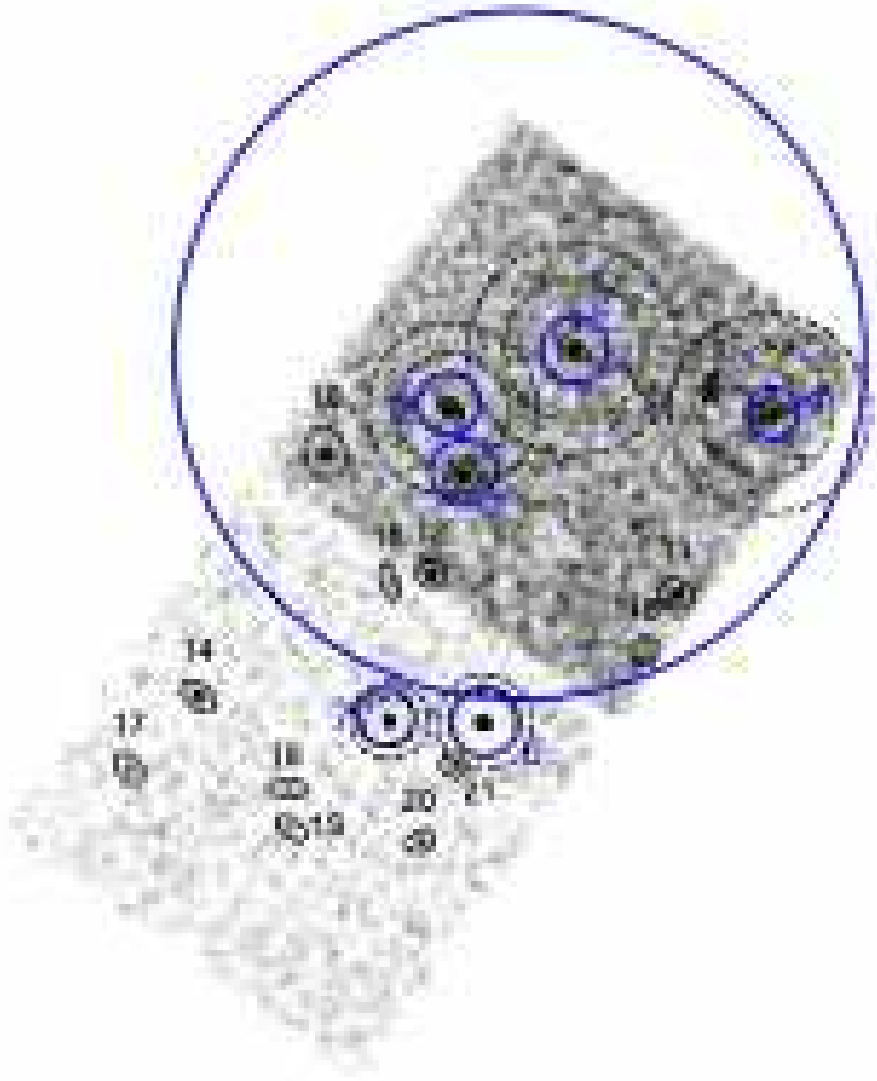}}}}
\caption{\cha\ image of G85.4+0.7 from the S2 and S3 chips, Gaussian smoothed with a radius of 5 pixels,
showing (in black) point source extraction regions and corresponding regions (in
blue) for \xmm\ data for 
comparison. The large circle indicates the diffuse region in the top panel of Figure~\ref{85.4reg} and the background regions are shown as dashed circles. (See the online version for colour figure.)\label{cha85.4}} 
\end{figure}

\begin{figure}
\scalebox{1.0}{\rotatebox{0}{\plotone{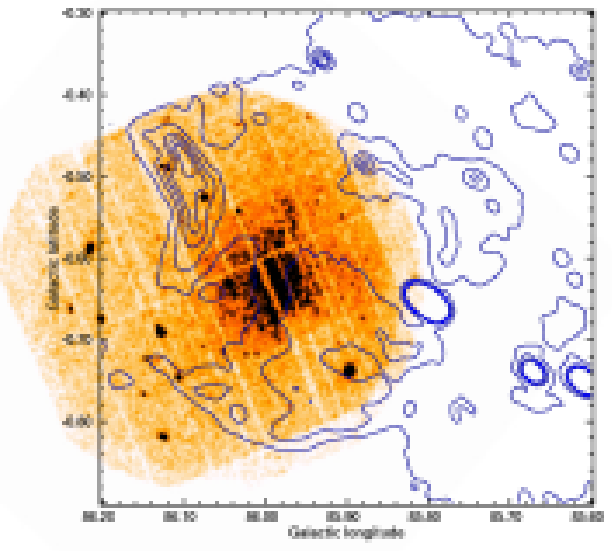}}}
\caption{0.5--2.5 keV \xmm\ X-ray mosaic image of G85.9$-$0.6 with the
  highest few radio contours
  overlaid in blue. The X-ray image is Gaussian smoothed with a radius of 3
  pixels. 
\label{85.9image}}
\end{figure}

\begin{figure}
\center{\scalebox{.45}{\rotatebox{0}{\plotone{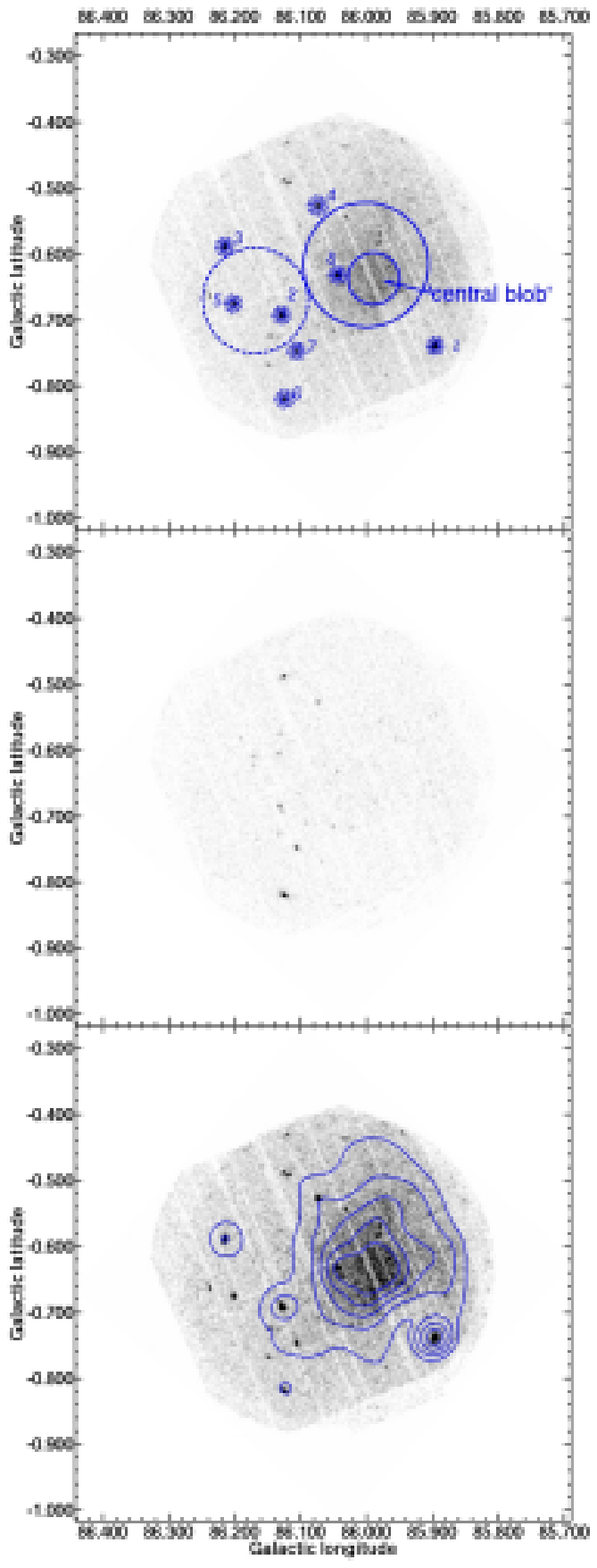}}}}
\caption{PN and MOS mosaic images of G85.9$-$0.6. Top panel: 0.5--2.5
  keV image, showing point
  sources (1-8) and the diffuse emission extraction region, as well as the region used for the central blob (see \S\ref{disc85.9}). The
  background region for the diffuse spectra is also shown
  as a dashed circle. The image is smoothed as in
  Figure~\ref{85.9image}. Middle panel: 2.5--10.0 keV image, similarly smoothed
  and with the same contrast as in the top panel. Bottom panel: 0.5--2.5
  keV image, Gaussian smoothed with a radius of 3 pixels and
  logarithmically scaled and adjusted in contrast
  to emphasize the diffuse emission and overlaid with
  contours of the 0.5--2.5 keV emission smoothed similarly to the \ro\
  contours in Figure~6 of
  \cite{kot01}. \label{85.9reg}} 
\end{figure}

\begin{figure}
\center{\scalebox{1.}{\rotatebox{0}{\plotone{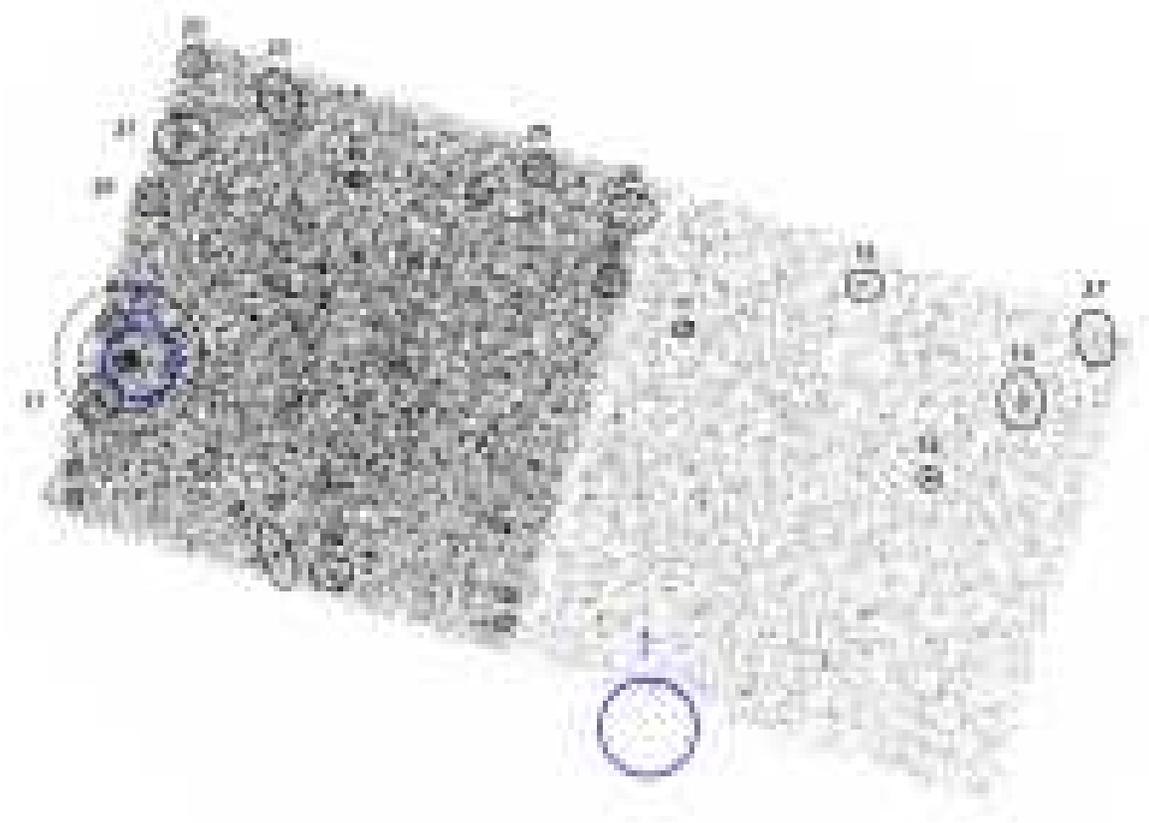}}}}
\caption{\cha\ image of G85.9$-$0.6 from the S2 and S3 chips, Gaussian smoothed with a radius of 5 pixels. 
Extraction regions for sources 1 and 8 from the \xmm\ image are shown as blue circles. (See the online version for colour figure.)\label{cha85.9}} 
\end{figure}

\begin{figure}
\center{\scalebox{.6}{\rotatebox{-90}{\plotone{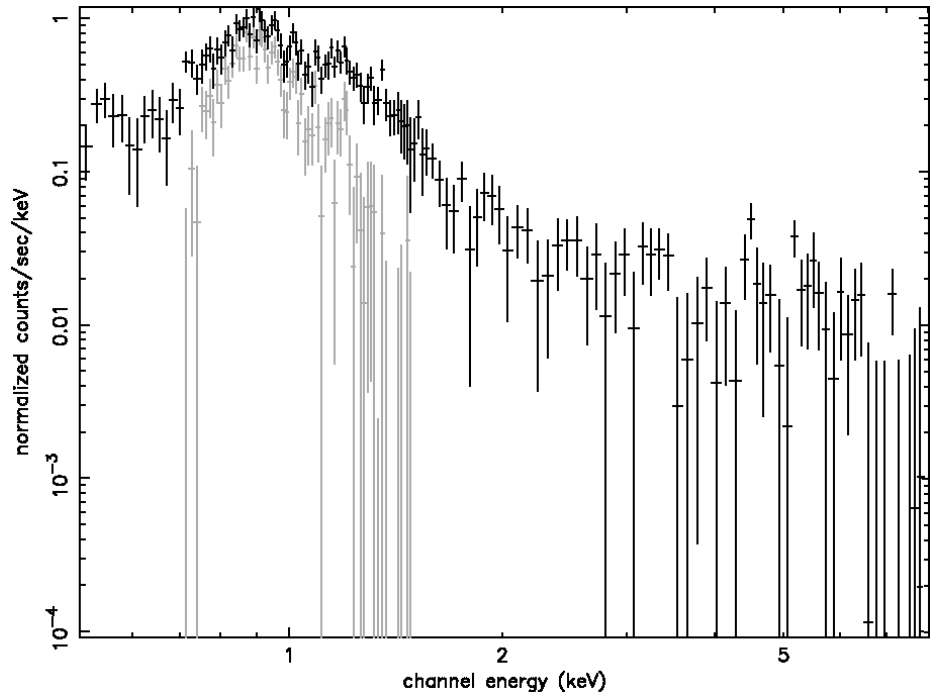}}}}
\caption{\xmm\ PN background subtracted 0.5$-$8.0 keV spectra of G85.4+0.7, using backgrounds 
from the observation (black) and from blank sky event files (grey). The
oversubtraction by the blank sky background is evident. The spectrum using the observation background is clearly background dominated above $\sim$2.5~keV. 
\label{PNback}} 
\end{figure}

\begin{figure}
\center{\scalebox{.6}{\rotatebox{-90}{\plotone{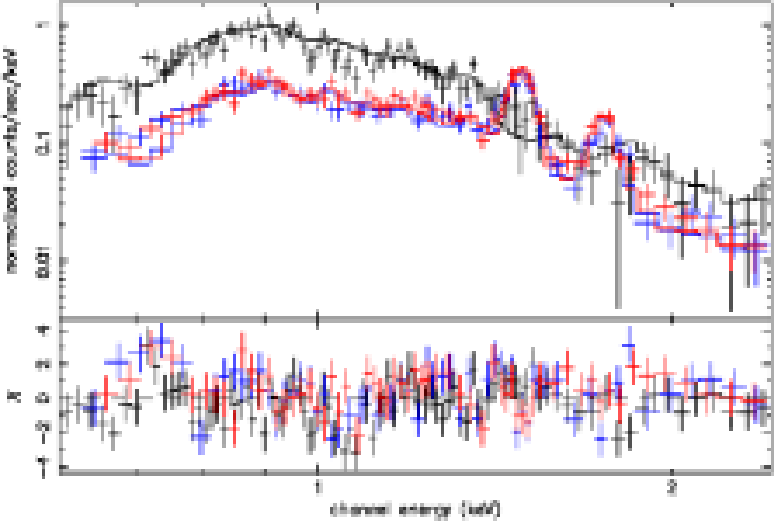}}}}
\caption{\xmm\ PN and MOS 0.5--2.5 keV spectra of G85.4+0.7, fit to an absorbed
  VPSHOCK. The background spectrum for PN was extracted in the same
  way as for the spectrum shown in Figure~\ref{85.4spec} and the
  background for the MOS instruments are extracted from the \xmm\ ESAS package described in \S\ref{bg}. The
  fitted parameters are given in the third column of Table~\ref{dfit}. The PN is shown
  in black, and the MOS 1 and 2 are shown in blue and red
  respectively. The bottom panel displays the residuals in units of
  $\sigma$.
Instrumental lines in the MOS spectra appear at 1.49 and
  1.75 keV (see \S\ref{bg} for details). The line above 1.75 keV is an
  elemental line which appears in spectra from all instruments.
\label{85.4esasspec}} 

\end{figure}

\begin{figure}
\center{\scalebox{0.6}{\rotatebox{-90}{\plotone{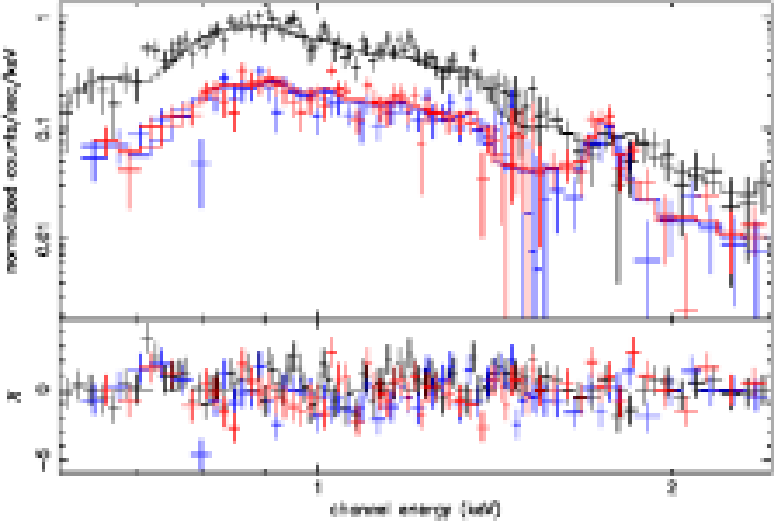}}}}
\caption{\xmm\ PN and MOS spectra of G85.4+0.7, fit to an absorbed VPSHOCK. The background spectra 
used here were extracted from the region shown in Figure~\ref{85.4reg}. The
  fitted parameters are given in the second column of Table~\ref{dfit}. The PN is shown
  in black, and the MOS 1 and 2 are shown in blue and red respectively. The bottom panel displays the residuals in units of $\sigma$. An instrumental line in the MOS spectra appears at 1.75 keV (see \S\ref{bg} for details).
\label{85.4spec}} 
\end{figure}

\begin{figure}
\center{\scalebox{0.6}{\rotatebox{-90}{\plotone{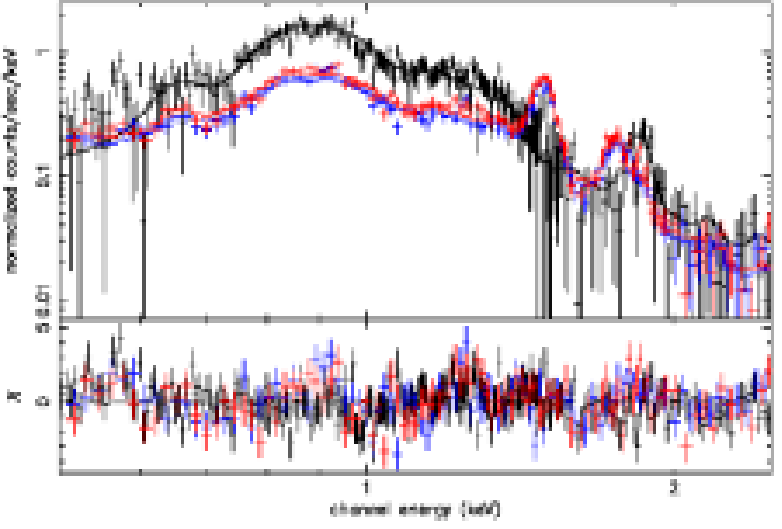}}}}
\caption{\xmm\ PN and MOS 0.5--2.5 keV spectra of G85.9$-$0.6, fit to an absorbed
  VPSHOCK. The background spectrum for PN was extracted in the same
  way as for the spectrum shown in Figure~\ref{85.9spec} and the
  background for the MOS instruments are extracted from the \xmm\ ESAS package described in \S\ref{bg}. The
  fitted parameters are given in the third column of Table~\ref{dfit}. The PN is shown
  in black, and the MOS 1 and 2 are shown in blue and red
  respectively. The bottom panel displays the residuals in units of
  $\sigma$.
Instrumental lines in the MOS spectra appear at 1.49 and
  1.75 keV (see \S\ref{bg} for details). The line above 1.75 keV is an
  elemental line which appears in spectra from all instruments.
\label{85.9esasspec}} 
\end{figure}

\begin{figure}
\center{\scalebox{0.6}{\rotatebox{-90}{\plotone{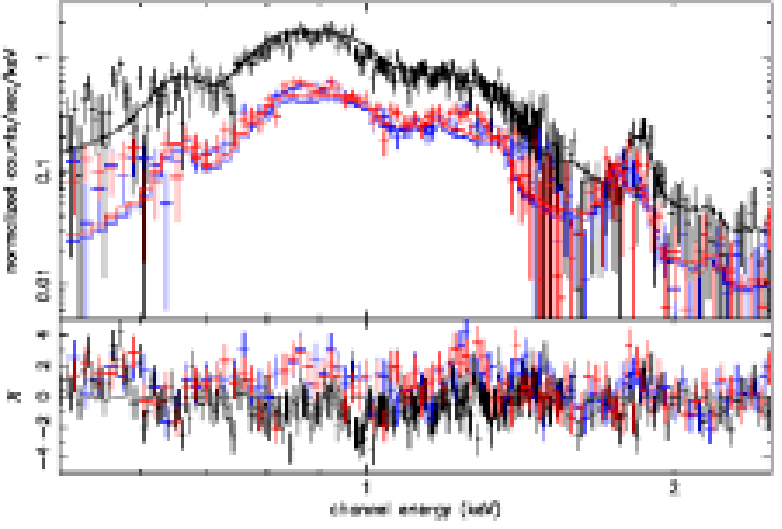}}}}
\caption{\xmm\ PN and MOS 0.5--2.5 keV spectra of G85.9$-$0.6, fit to an absorbed VPSHOCK. The background spectra 
used here were extracted from the region shown in
Figure~\ref{85.9reg}. The
  fitted parameters are given in the fourth column of Table~\ref{dfit}. The PN is shown
  in black, and the MOS 1 and 2 are shown in blue and red respectively. The bottom panel displays the residuals in units of $\sigma$. An instrumental line in the MOS spectra appears at 1.75 keV (see \S\ref{bg} for details). The excess counts in the MOS spectra around 1.4 and 0.85 keV are most likely due to instrumental effects.
\label{85.9spec}}
\end{figure}

The previous \ro\ data of the SNR regions produced contours of the
diffuse emission but did not resolve any point sources. The presently considered
\xmm\ and \cha\ data allow for point sources to be resolved and located within
the field of view. Nine clearly distinguishable point sources are seen
on the image of G85.4+0.7 (Figure~\ref{85.4reg}), of which six (sources 1, 2, 3, 4, 6 and 7) 
are in common with 
point sources in the \cha\ observation, as shown in Figure~\ref{cha85.4}. Twelve additional
point sources are clearly resolved in the \cha\ observation (labeled as sources 10 
through 21 in Figure~\ref{cha85.4}), but they do not possess
sufficient counts to allow for meaningful spectral analysis to be done. It should be noted that 
the B stars in \cite{kot01} are outside of the field of view of the X-ray observations.
Eight point sources are seen on
the \xmm\ image of G85.9$-$0.6 (Figure~\ref{85.9reg}) of which only one source (source 8) is visible on 
the \cha\ image, which is pointed toward the region encompassing sources 1 and 8 in 
Figure~\ref{85.9reg}. Figure \ref{cha85.9} also shows 19 additional point sources from the \cha\ observation 
(labeled 9 through 27), but none of them possess sufficient counts for their spectral parameters to be sufficiently constrained.
In the present work,
the nine point sources detected with \xmm\ in the field of G85.4+0.7 and eight in the field of G85.9$-$0.6 are analysed using spectral and timing
techniques, to locate any candidates for identification as neutron stars or pulsars which
would have formed at the time of the supernova explosion.

\subsection{Source Identification}\label{sid}

To identify the X-ray point sources in the G85.4+0.7 and G85.9$-$0.6
fields, {\it ewavelet}, a wavelet detection algorithm which is part of
the SAS 6.5 package, is used. For each source, the output of the
routine gives position on the image and in sky coordinates, source
counts, source extent, as well as errors in those quantities. Because the PN
and MOS instruments have slightly different fields of view, only those
sources found in the combined image which are also found on the PN
image are analysed. The sources must also have an extent (size on the
image) similar to the PSF at that location on the image, and they 
must contain enough counts ($>50$) for the spectral and timing analysis. 

A catalogue of newly discovered X-ray point sources is given in
Tables~\ref{pscat85.4} and \ref{pscat85.9}. The letters XMMU in the object designations
indicate that they were discovered in \xmm\ data and a prefix of CXO indicates a discovery in \cha\ data. The \ro\ All Sky
Survey catalogue was checked to see if any of the sources appear there,
and source 1 in the G85.9$-$0.6 image (Figure~\ref{85.9reg}) is well
within the 24$^{\prime\prime}$ $1\sigma$ error circle of the
coordinates of the \ro\ All Sky Survey object 1RXS J205911.3+444730,
and source 1 in the G85.4+0.7 image (Figure~\ref{85.4reg}) lies
marginally within
the 24$^{\prime\prime}$ $1\sigma$ error circle of the RASS object 1RXS J205058.7+452135. 
No
other point sources in this study match any in the RASS catalogue,
indicating that these point sources are too faint to be
included in the RASS catalogue.

The sky coordinates of sources meeting the above criteria are searched within
the extent, which is approximately the PSF size, in various
catalogues, including the USNO-A2.0 catalogue \citep{mon98}, the
USNO-B1.0 catalogue \citep{mon03}, the SKY2000
catalogue \citep{my02}, the catalogue given in \cite{gua92}, and the 2MASS
catalogue \citep{2mass}, to determine whether the sources have already been
identified in another waveband. Some of these catalogues 
give blue magnitude and if this is converted into flux with the
relation $m_1-m_2=-2.5 log(F_1/F_2)$ and compared with the X-ray flux, 
this ratio is one factor that can be checked to favour identification
as a neutron star. In the above
equation, $m_1$ and $m_2$ are the blue magnitudes of the object and a
standard star and $F_1$ and $F_2$ are their fluxes. Typical neutron
stars have a flux ratio in optical to X-ray bands of $\sim 10$ \citep{lyn98},
whereas X-ray emitting O- or B-type stars such as $\eta$ Carinae \citep{cor00} or $\tau$
Scorpii \citep{mew03} have a ratio of $\sim 10^7$, which enables an identification to
be made of neutron star candidates. The optical to X-ray flux ratios
for the point sources, matched with each of the catalogue objects within
their extents are shown in Tables~\ref{pscat85.4} and \ref{pscat85.9}. These values represent
the optical to X-ray flux ratio should the catalogue object match the
X-ray source. If the optical catalogue source and the X-ray point source
are not the same object, the ratio is meaningless. However, an X-ray
source without any optical counterparts within the PSF, such as
source 6 or 7 from the G85.9$-$0.6 data, may be a good neutron star
candidate. In other words, this test does not exclude X-ray point
sources as neutron star candidates but rather indicates that a source
emits in X-rays, and is much fainter in the optical waveband, and is
unlikely to be a stellar object. The optical to X-ray ratios for all
point sources for which the ratio can be calculated (ie all sources
except 6 and 7 in the G85.9$-$0.6 field) is $>10^{11}$,
which is much greater than the ratio expected for neutron stars,
indicating that these objects are not neutron stars,
assuming that the optical sources are the true counterparts. Other potentially 
interesting objects (e.g. neutron star or AGN candidates) are \cha\ sources 14, 16, 17, 18, 19, 21, 24, 25, and 27 in the 
G85.9$-$0.6 field, three of which have no counterpart at all (21, 24, and 27), and the 
rest of which have either only a 2MASS counterpart or a 2MASS counterpart which is 
a much better match for the position than the optical counterpart. These objects are 
shown in Figure~\ref{cha85.9} and listed in Table~\ref{pscat85.9}.

\subsection{Spectral Analysis}\label{psspec}
\begin{figure}
\center{\scalebox{0.6}{\rotatebox{-90}{\plotone{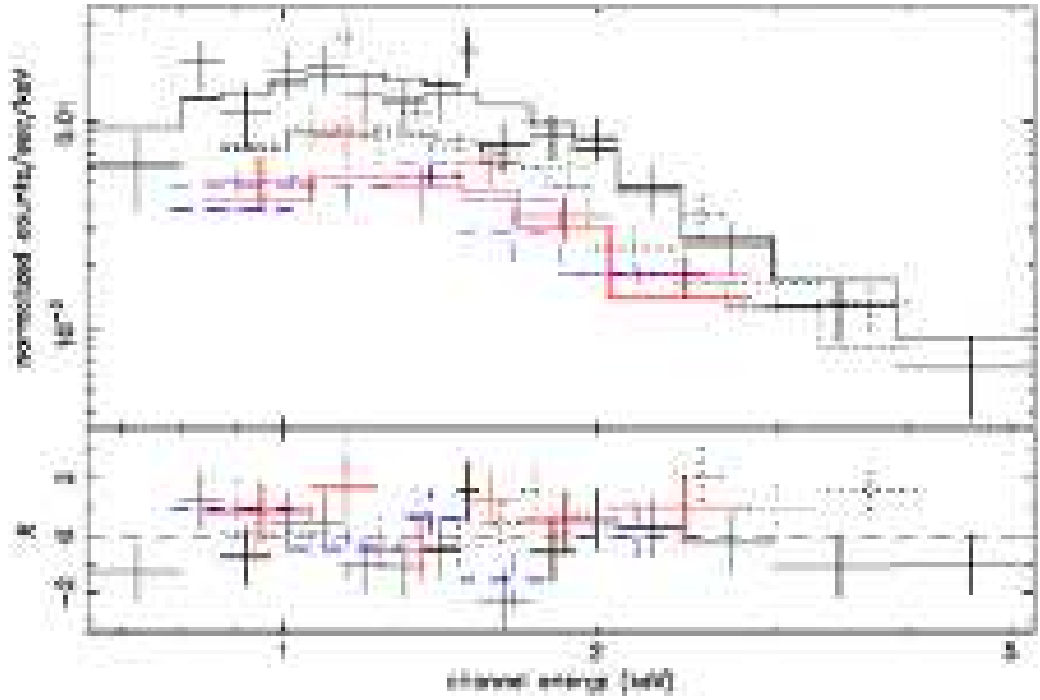}}}}
\caption{\xmm\ PN, MOS, and \cha\ 0.5--5 keV spectra of G85.4+0.7 source 1, fit to an absorbed power law. The
  fitted parameters are given in Table~\ref{psspectab}. The PN is shown
  in solid black, the MOS 1 and 2 are shown in dashed blue and solid red respectively, 
and the \cha\ spectrum is shown as dotted black. The bottom panel displays the residuals in units of $\sigma$.
\label{85.4psspec}} 
\end{figure}

In addition to the optical to X-ray flux comparison described in
\S\ref{sid}, the X-ray spectra of the point sources can be examined to
determine the likelihood that any of them are neutron stars. The X-ray
spectrum of a rotation powered pulsar is typically hard with a
power law photon index ($\Gamma$) of approximately 0.5$-$2.1 (e.g. \cite{gott06}),  though there may be an
additional blackbody component from the neutron star surface which
could dominate the spectrum, depending on the age and the star's
magnetic field. Anomalous X-ray pulsars typically have photon indices between
2.4 and 4.6 (e.g. \cite{wood04}).
The
background-subtracted spectra of the point sources, where the
background region consists of an annulus
surrounding each point source, are fit to an absorbed power law and the resulting
fitted parameters are shown in Table~\ref{psspectab}. In addition, the
0.5--2.0 keV and 2.0--10.0 keV counts are given, which is particularly
useful for estimating an X-ray hardness ratio when the fit to an
absorbed power law yielded a large $\chi^2$ value, indicating an
unsuitable model. Sources 1, 4, and 9 in the G85.4+0.7 field and sources 4,
6 and 7 in the field of G85.9$-$0.6 have been identified as neutron star
candidates. Sources 1 and 9 of G85.4+0.7 both have photon indices of
$\sim 2.5$, which does not rule them out as neutron star candidates
but they both have a high optical/X-ray flux ratio.  Both source 4 of
G85.4+0.7 and source 4 of G85.9$-$0.6 have a relatively soft X-ray
spectrum compared with typical pulsars (and the $\chi^2$ value for the
fit of source 4 on G85.9$-$0.6 indicates that the power law model does
not match the data well) and both have a high optical/X-ray flux 
ratio. However, it is not certain that the optical sources associated with 
these objects are the true counterparts. Neutron star candidates 6 and
7 in the G85.9$-$0.6 field have a photon index typical to neutron stars and no 
optical counterpart. However, source 6 has a column density ($N_{\rm H}$) which
is greater than that of the SNR itself, which indicates, along with
its position relative to the SNR, that it is unlikely to be the
associated neutron star. Source 7 has a $N_{\rm H}$ value which is
similar to that of the SNR, but its position indicates that it is
probably not associated with the SNR. When considered separately, \xmm\ and \cha\ spectra of 
G85.4+0.7 point sources 1 and 3 yield 
spectral parameters which agree with each other well within the range of uncertainty. The \cha\ spectra 
of sources 2, 4, 6, and 7 do not contain enough counts for independent spectral 
analysis but the \cha\ spectra for these sources are included in the
analysis leading to entries in the appropriate 
rows of Table~\ref{psspectab}, and the spectral parameters of the combined \xmm\ and \cha\ 
spectra of these sources lie within the uncertainty of those for the \xmm\ spectra alone. 
As an example, the \xmm\ PN and MOS and \cha\ spectra of G85.4+0.7 source 1 is shown along
with a fit to an absorbed power law in Figure~\ref{85.4psspec}.

\subsection{Timing Analysis}
A timing analysis is done on the PN data for the neutron star
candidates identified above to search for pulsations. The timing
resolution in PN full
window mode ($68.7$ ms) for both G85.4+0.7 and G85.9$-$0.6 allows for a
search up to $\sim 7$ Hz, which would fail to identify fast 
rotation-powered
pulsars, but would identify slowly rotating anomalous X-ray pulsars. The PN events file is first
barycentre corrected. The photon arrival times for a region
surrounding each source are used first in a fast Fourier transform
(FFT) search to identify possible frequencies that should be
investigated further. Around each frequency identified by the FFT search,
Rayleigh \citep{lea83} (also known as $Z_1^2$), $Z_2^2$ and $Z_3^2$
\citep{buc83}, and epoch folding searches are done. The statistical
significances of any identified frequencies are calculated using the
$\chi^2$ probability along with the number of degrees of freedom,
which is $2n$ for the $Z_n^2$ search and $n_{\rm bins}-1$ for the
epoch folding search, where
$n_{\rm bins}$ is the number of bins in the lightcurve. In 
this case the number of bins used is 12 and 20 for the two epoch folding tests performed on 
the data, chosen because those numbers of bins produce a relatively
detailed lightcurve while maintaining a reasonable number of counts per bin. 
With regard to the $Z_n^2$ test, the $Z_1^2$, $Z_2^2$, and
$Z_3^2$ searches are performed and would identify typical pulsar X-ray
lightcurves, which usually exhibit either a broad variation or two or
three peaks per cycle.

The timing
resolutions of the MOS instruments (which, in the data acquisition mode used
for these observations, is 2.6 seconds) and \cha\
(3.24 seconds) are not sufficient for
meaningful timing analysis of this type to be done. 

Of the six neutron star candidates found from the point source
spectra of the two observations, none were found to exhibit a periodic
signal. However, a future dedicated timing search may uncover one of
these sources as a pulsar. 

\subsection{Results of point source analysis}
Based on a combination of the optical to X-ray ratios (assuming the optical sources are the counterparts of the X-ray sources), distances from
the SNR centres, spectral parameters, and comparisons of $N_{\rm H}$ value
to that of the diffuse emission, none of the point sources is likely
to be the neutron star candidate associated with the SNR. However, further observations of 
the \cha\ objects may reveal one of them to be a neutron star or AGN.

\section{Distances}\label{dist}
It is interesting to note that the absorbing \ion{H}{1} column density derived
from the X-ray spectra is smaller for G85.9$-$0.6 than it is for G85.4+0.7,
even though G85.9$-$0.6 was predicted to be further away by \cite{kot01}.
The knowledge about the foreground \ion{H}{1} column density gives us another
option to constrain the distance to these objects. Both SNRs are believed to be
located outside the solar circle. In the outer Galaxy the radial velocity
is decreasing monotonically with distance, independent of the Galactic rotation
model used. Hence, we can integrate foreground atomic and molecular hydrogen
as a function of distance, by integrating the spectroscopic data down to
the appropriate radial velocity. Neutral hydrogen data are from the Canadian
Galactic Plane Survey \citep[see][for further information]{tay03}, and
the $^{12}$CO(1-0) molecular line data are from the Columbia CO survey of
\citet{dame87}.

Since atomic hydrogen is usually optically thick we could not simply
integrate the \ion{H}{1} emission, as emission does not represent all of the hydrogen actually
present along a Galactic line of sight. To correct for this we have to determine
the optical depth of the \ion{H}{1} along the line of sight. One way of doing
this is looking at background point sources, preferably of extragalactic
origin, which are very bright radio continuum emitters at 1420~MHz. If these
sources are bright enough we can see their emission being absorbed by the
foreground \ion{H}{1} and use them to probe the ISM along their line of sight
through the whole Galaxy. The absorbing neutral hydrogen column density
integrated over the velocity interval dv is then defined by $N_{\textrm{\ion{H}{1}}}(v)
[cm^{-2}] = 1.8224\times 10^{18}\, T_s(v)\, \tau(v)\, dv$. $\tau$ is the
optical depth, which can be determined from the absorption profile by
$\tau(v) = -ln((T_{on}(v) - T_B(v)/T_{bg}) + 1)$, here $T_{bg}$
is the brightness temperature of the absorbed background 
source, $T_{on}$ is the continuum subtracted \ion{H}{1} brightness temperature
at the position of that source, and $T_B$ is the brightness temperature
of the absorbing \ion{H}{1} cloud. $T_B$ is represented by the average off-source
\ion{H}{1} brightness temperature, which is determined in an $1\arcmin$ wide elliptical
annulus just outside the absorbed background source. $T_s$ is the spin
temperature of the absorbing cloud, which is 
defined by $T_s(v) = T_B(v)/(1 - e^{-\tau(v)})$. Equations for $N_{\textrm{\ion{H}{1}}}(v)$,
$\tau(v)$, and $T_s(v)$ can be found in e.g. \cite{roh04}.

To derive the complete foreground \ion{H}{1} column we have to add
twice 
the molecular hydrogen column density $N_{H_2}$. This is derived
over the velocity interval $dv$ using its relation to the CO brightness
temperature $T_B^{CO}$, determined by \cite{dame01}: $N_{H_2}(v) [cm^{-2}] = 1.8\times 10^{20}\,T_B^{CO}(v)\,dv$.

\begin{figure}
\center{\scalebox{1.0}{\rotatebox{0}{\plottwo{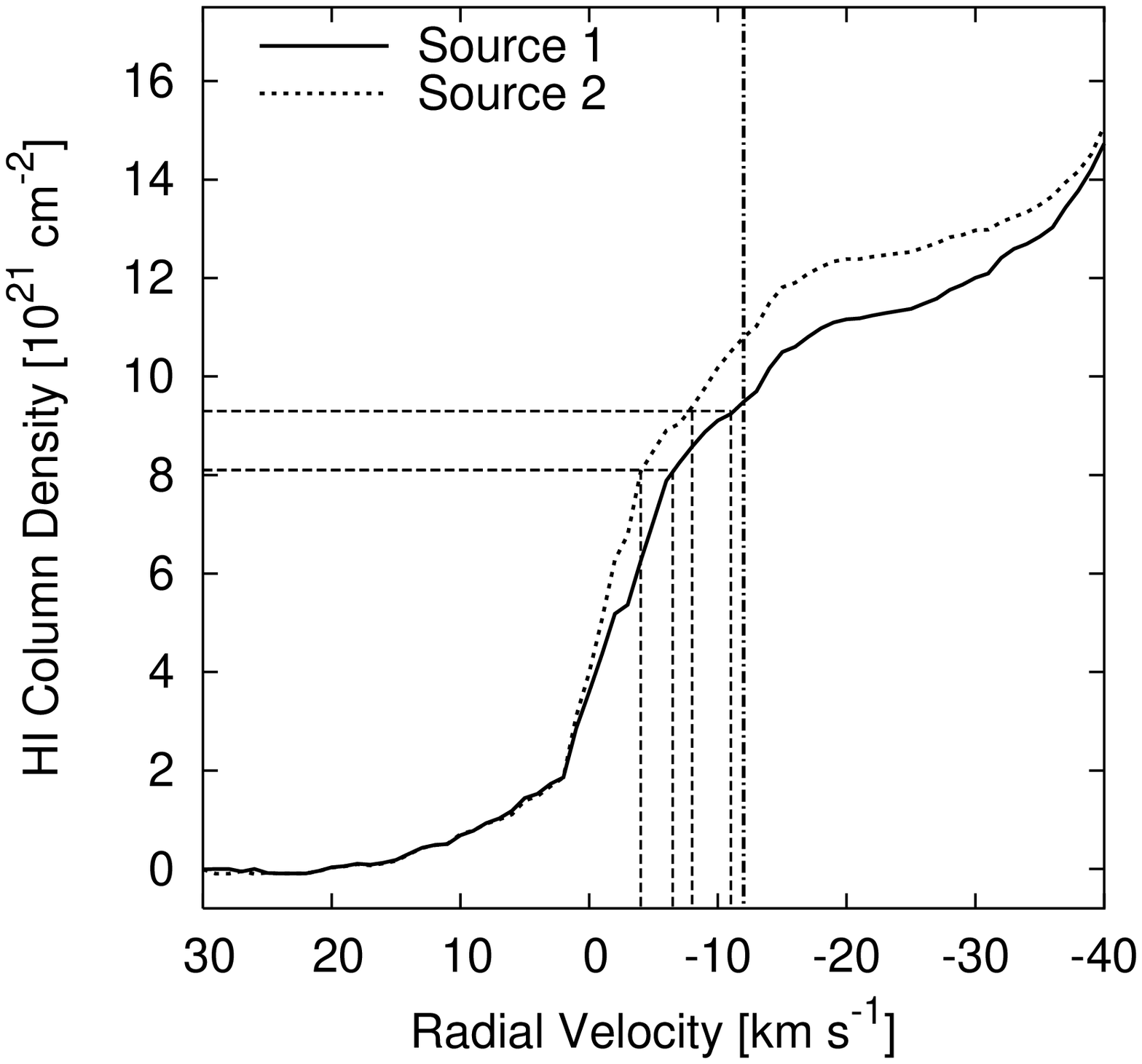}{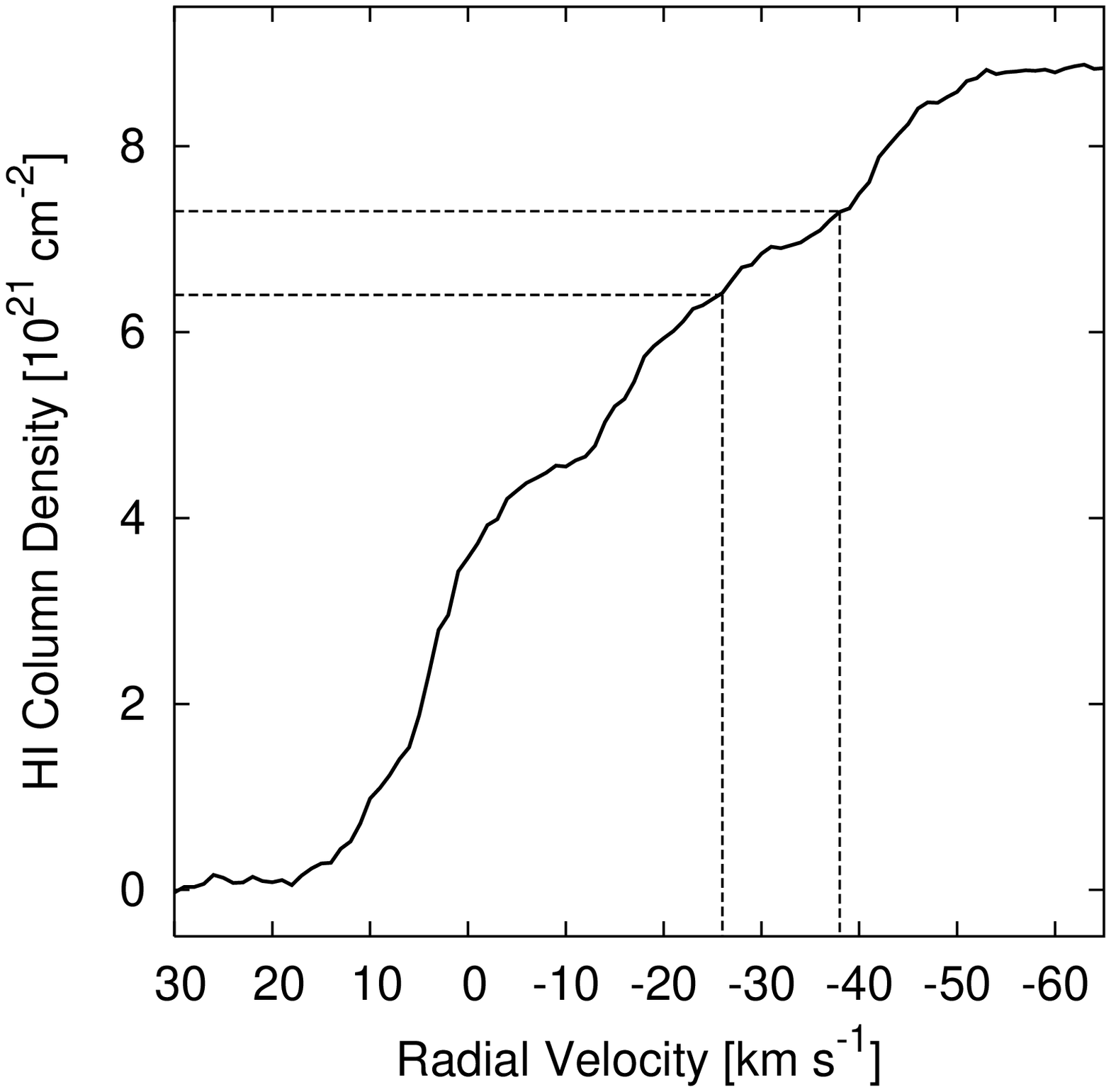}}}}
\caption{Integrated foreground hydrogen column density profiles as a function
of radial velocity for the two SNRs G85.4+0.7 (left: solid and dotted line)
and G85.9$-$0.6 (right: solid line). The dashed lines represent the upper and
lower limits for the absorbing \ion{H}{1} column density determined from the X-ray
spectra. The dotted-dashed line in the diagram for G85.4+0.7 represents the
radial velocity of the related stellar wind bubble determined by Kothes et al.
(2001). 
\label{hicol}}
\end{figure}

For the SNR G85.4+0.7 we used the absorption profiles of two
background sources to determine the foreground atomic hydrogen
column density. One source is located at the very centre of
the remnant (Figure~\ref{hicol}: source 1) at $\ell = 85.31\degr$ and
$b = +0.66\degr$ and the other just to the south
(Figure~\ref{hicol}: source 2) at $\ell = 85.31\degr$ and $b = +0.32\degr$.
For G85.9$-$0.6 we used the bright source at $\ell = 85.80\degr$ and
$b = -0.65\degr$. To calculate the amount of foreground molecular
hydrogen we averaged the Dame et al. (2001) CO data, which has a
pixel size of $30\arcmin$ over the four pixels closest to the centre
of the X-ray emission. The final combined \ion{H}{1} column density
profiles are displayed in Figure~\ref{hicol}.

If we now compare the absorbing \ion{H}{1} column density, which
we derived from the X-ray spectra (see Table \ref{dfit}), with the
$N_{\textrm{\ion{H}{1}}}$ - velocity diagrams in Figure~\ref{hicol}, we can get
an estimate for the radial velocities of the SNRs.
For the SNR G85.4+0.7 we derive a radial velocity of about
$-9$~km s$^{-1}$ averaging over source 1 and 2. This nicely
confirms the radial velocity of $-12 $~km s$^{-1}$, which was
determined for the stellar wind bubble surrounding G85.4+0.7
by Kothes et al. (2001). For G85.9$-$0.6 no previous estimate
of the radial velocity exists. The comparison of the absorbing
\ion{H}{1} column (Table~\ref{psspectab}) with the $N_{\textrm{\ion{H}{1}}}$ - velocity diagram
in Figure~\ref{hicol} results in a radial velocity estimate of
$-32\pm 6$~km s$^{-1}$.

Previously, \citet{kot01} found a distance to G85.4+0.7 of about
3.8$\pm$0.6~kpc, based on the radial velocity
of its host stellar wind bubble using a flat rotation model for the
Galaxy with a Galacto-centric radius of
8.5~kpc for the Sun at a velocity of 220~km s$^{-1}$ around the
Galactic centre. A distance of $\sim$5~kpc
was predicted for G85.9$-$0.6. The radial velocities of G85.4+0.7
($-$12$\pm$3~km s$^{-1}$) \& G85.9$-$0.6 ($-$32$\pm$6~km s$^{-1}$) indicate
they are beyond the Solar circle \citep[$R_{0}=$7.6~kpc,][]{eise05}, but from
$T_{b}\left(\ell,~v\right)$ diagrams in this direction they are
not residents of the Perseus Spiral arm (which shows as a large \ion{H}{1}
feature extended in longitude, very nearly centred on $-$40~km s$^{-1}$).

A new kinematic-based distance method has been developed by
\citet{fmac06}. The approach is based on a model of the Galactic
\ion{H}{1} density distribution and velocity field, that is
\textit{fitted} to observations, rather than relying on a purely circular
rotation model assigned to the object (as in standard kinematics). The
model's density component is that of a warped thick disk of \ion{H}{1},
with axisymmetric features, and a two-arm density wave pattern in the
disk. The velocity field component models a non-linear response of the gas
to the density wave \citep[see][]{robe69, wiel79}, producing shocks at the
leading edge of major arms, and streaming motions within the disk,
depending on the location of an object in the spiral phase
pattern. This distance method has been shown
to accurately reproduce spectrophotometric distances to \ion{H}{2}
regions throughout the second quadrant of the Galactic plane.

\begin{figure}
\center{\scalebox{0.6}{\rotatebox{0}{\plotone{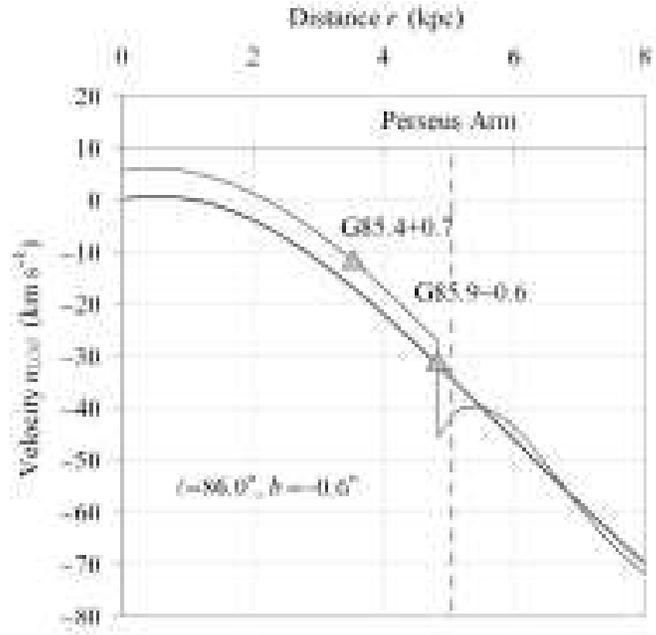}}}}
\caption{The velocity field of the \ion{H}{1} gas
towards $\ell=$86.0$\degr,~b=-$0.6$\degr$, from the \ion{H}{1} modelling
method of \citet{fmac06}. SNRs G85.4+0.7 \& G85.9$-$0.7 are marked with
triangles showing their respective distances of 3.5 and 4.8 kpc. The
Perseus Spiral arm's peak density is at 5.0 kpc and the spiral shock that
precedes the arm is at 4.8~kpc. The solid dark line is the underlying
circular velocity field fitted in this direction, and is a power-law in
galactocentric distance \citep[see][]{fmac06} with a gradient similar to a
Schmidt rotation curve.
\label{vfield}} 
\end{figure}

The best fit synthetic \ion{H}{1} profile (with velocity field as in
Figure \ref{vfield}) towards these objects shows that G85.4+0.7 is
3.5$\pm$1.0~kpc distant. This distance indicates G85.4+0.7 is within the
Local spiral arm, a feature in between the major Sagittarius and Perseus
spiral arms of the Milky Way. While the velocity of G85.9$-$0.6 is less
certain compared to G85.4+0.7 \citep[determined by association with
\ion{H}{1}; see][]{kot01}, it should be noted that in G85.4+0.7's case,
comparison of the X-ray absorbing column to the total integrated hydrogen
nuclei (in \ion{H}{1} and $^{12}$CO data) gives a good velocity estimate.
Hence, we can be reasonably sure that $-$32~km s$^{-1}$ is near to the
true velocity of G85.9$-$0.6 as well. This places it within the Perseus
arm spiral shock, which is 4.8$\pm$1.6~kpc in this direction (the arm's
potential minimum is 5.0~kpc). For any velocity in the range $-$40$\leq
v_{LSR} \leq -$18~km s$^{-1}$, the distance range shown in the fitted
velocity field (Figure \ref{vfield}) is small, about 4.1$-$4.8~kpc.

Although SNRs stemming from Type II events are primarily found within the
arms \citep[as massive progenitors are born mainly in the arm's
shock,][]{wiel79}, the location of G85.9$-$0.6's within the shock does not
necessarily vitiate its identification as a Type Ia. For example, Tycho's
SNR is known to be Type Ia, but its velocity and distance clearly
associate it with the Perseus Spiral arm. It is possible that the binary
precursor of G85.9$-$0.6 migrated into the arm along its Galactic orbit
before exploding as a Type Ia event.

\section{Discussion}\label{disc}

\subsection{Determination of shock age and mass of emitting gas}

To estimate $t$, the shock age, for the
SNRs, the relation $t=\tau/n_{\rm e}$ is used, where $\tau$ is the
upper limit of the ionization timescale from the VPSHOCK model. The electron density
$n_{\rm e}$ is determined from the distance in cm $D_{\rm cm}$, angular
size in radians $\alpha$, given by the diameter of the extraction region, and the
relation $n_{\rm H}=n_{\rm e}/1.2$, where $n_{\rm H}$
is the volume density of hydrogen within. Given
that ${\rm Norm}=\frac{\int n_{\rm e}n_{\rm{H}}dV}{10^{14}(4\pi D_{\rm cm}^2)}$, where
Norm is the normalization of the
VPSHOCK model, which is proportional to the emission measure, and the
electron and hydrogen densities $n_{\rm e}$ and $n_{\rm{H}}$ are
assumed to be uniform, a volume filling factor $f$ is employed so that
$n_{\rm e}=(\frac{2.88\times 10^{15} ({\rm Norm})}{f \alpha^3D_{\rm cm}})^{1/2}$. From
these calculations, it is determined that for G85.4+0.7, $n_e \approx
(0.11\pm0.03)(fD_{3.5})^{-1/2}$ for the ESAS background (for the MOS
instrument) or $(0.11^{+0.05}_{-0.04})(fD_{3.5})^{-1/2}$
 for the observation background, and $t =(18^{+17}_{-9})(fD_{3.5})^{1/2}$
 kyr for the ESAS background or $(23^{+26}_{-14})(fD_{3.5})^{1/2}$
 kyr for the observation
background. For G85.9$-$0.6, $n_e \approx (0.07^{+0.03}_{-0.02})(fD_{4.8})^{-1/2}$ for the
ESAS background or $(0.07\pm0.03)(fD_{4.8})^{-1/2}$ for the observation
background and $t = (30^{+12}_{-13})(fD_{4.8})^{1/2}$ kyr for the ESAS
background or $(22^{+11}_{-10}) (fD_{4.8})^{1/2}$ kyr for the observation
background. These are larger than the age estimates from the radio
data.

The mass of the emitting gas is
calculated based on spherical emitting regions with the size given by
the extraction radius (given in Table~\ref{dfit}), composed of
92\% hydrogen and 8\% helium, and the relation $n_{\rm H}=n_{\rm
  e}/1.2$ is used as for the above calculation of $t$. The mass of the
emitting gas in G85.4+0.7 is
$(2.8\pm1.5) D_{3.5}^{5/2} M_{\sun}$ for the ESAS
background and $(2.8^{+1.8}_{-1.6})D_{3.5}^{5/2} M_{\sun}$ for the
observation background and that for G85.9$-$0.6 is
$(2.7^{+1.8}_{-1.6}) D_{4.8}^{5/2} M_{\sun}$ for the ESAS
background and $(2.7\pm1.8)D_{4.8}^{5/2} M_{\sun}$ for the
observation background.

\subsection{Background subtraction}

The background subtraction was problematic, particularly for the
spectrum of G85.4+0.7, which is fainter and required a different model
to be fit to it depending on which background region on the
observation was used, leading to a very careful selection of the
background extraction region, the process of which is described in
\S\ref{bg}. Instrumental lines, which are fortunately different for PN
and 
MOS, nevertheless increased the value of $\chi^2$ for the combined fits, even
when the background was chosen very carefully, and needed to be
explicitly fit when the ESAS background was used, as described in \S\ref{bg}. In addition, the soft
spectra of the SNRs exhibit large amounts of noise in the
high energy end of the spectra, allowing fits to be made only up to
2.5 keV. Future longer observations of these SNRs will
help to resolve these difficulties, allow for better fits to the
abundances, eliminate ambiguities in the spectral results, and perhaps
allow for conclusive identifications of neutron star and 
pulsar candidates in addition to other point sources.

\subsection{G85.4+0.7}\label{disc85.4}

\begin{figure}
\scalebox{1.0}{\rotatebox{0}{\plotone{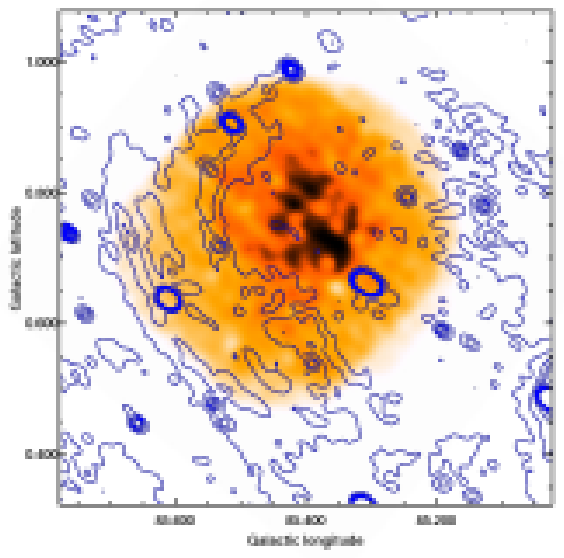}}}
\caption{0.5--2.0 keV \xmm\ X-ray mosaic image of G85.4+0.7 with radio
  contours overlaid in blue. The X-ray image has had the point sources
  removed and is Gaussian smoothed with a radius
  of 1$^\prime$ to match the radio resolution, in order to compare the radio
  and diffuse X-ray emission regions. The radio contours were chosen to approximate the
  appearance of the top image in Figure~2 in \cite{kot01}.
\label{85.4imagens}}
\end{figure}

A comparison between the morphologies of the X-ray diffuse emission and the radio emission
of G85.4+0.7 can be seen in Figure~\ref{85.4imagens} in which the
X-ray point sources have been removed and the X-ray image has been
smoothed to 1$^{\prime}$ to match the radio contours. The diffuse
X-ray emission exhibits some structure, and it lies in the approximate
centre of the radio shell. A spectrum was extracted from the central
blob, the position and size of which is shown in the top panel of Figure~\ref{85.4reg}, and it could not be adequately fit with a non-thermal power law
model, indicating that a pulsar wind nebula origin for the emission can
be ruled out. 

The fact that the SNR does not exhibit any limb brightening and the X-ray
emission is mostly concentrated in the centre, as projected in
the plane of the sky and three dimensionally, can be interpreted as
evidence that the X-ray emission is produced by the ejecta. The fact
that there appears to be no emission from the swept up material can be
explained if the remnant is evolutionarily young. 

With the ejecta
interpretation, if it is assumed that the free electrons are evenly
distributed in the extraction area, the resulting ages are 18 and 23
kyr for the two backgrounds. In Figure~\ref{85.4imagens} it can be
seen that the outer radio shell has an approximately constant radius,
whereas the inner shell has a smaller radius in the vertical centre,
which expands as the latitude changes, and this indicates that the SNR is
moving either toward us or away from us. The two shells appear to
meet near the bottom of the image, so the radius of the SNR can be
approximated as that of the outer shell, which is $\sim0.3^\circ$ on
the image. The average velocity of the
expanding SNR is then 910 or 730 km/s for the two ages. However, the
morphology of the X-ray emission is not very smooth, indicating that
the filling factor $f<<1$. Assuming $f=0.2$, the electron density $n_e$
would be 0.25 cm$^{-3}$ for either the ESAS and observation background, 
and the ejecta masses would be 1.3 $M_{\sun}$ for
either background. The shock age would then be 8000 or 10300 years,
and the average expansion velocity would be 2150 or 1640 km/s. For a
$10^{51}$ erg supernova explosion, the ejecta mass would be 22 or 37
$M_{\sun}$, or 2.2 or 3.7 $M_{\sun}$ for a $10^{50}$ erg
explosion, indicating that the lower energy explosion would result in
an ejecta mass consistent with a young freely expanding SNR.

The radio continuum emission indicates that there is some swept up
material. Assuming that the density inside the stellar wind bubble was
$\sim 0.01$ cm$^{-3}$ before the explosion, the SNR would have swept up
$\sim 6.5$ $M_{\sun}$ of material, which is on the order of the
ejecta mass, and means that the SNR is in the transition between free
expansion and Sedov expansion.

The slightly above-solar abundance of O in the diffuse spectra reinforces the hypothesis that
the supernova most likely resulted from a core collapse, though the
large error bars weaken the argument. The abundances of all elements
other than O and Fe are below solar, but they would be enhanced if the
X-rays were from the ejecta. However, it is possible that the spectral
parameters for the abundances are affected by the poor quality of the
spectra and by the uncertainties associated with the background
subtraction. 

Since the most likely origin of G85.4+0.7 was a core
collapse supernova (see \S\ref{dist}), it is possible that a
neutron star or pulsar 
which is associated with this SNR can be found. Given its power law X-ray 
spectrum, proximity to the centre of the remnant, and
similar $N_{\rm H}$ value to the diffuse emission, source 1 in Figure
\ref{85.4reg} is possibly the associated neutron star, but sources 4
and 9 are
within the radio shell and are also candidates, given their
spectral parameters. However, the low fit quality of source 4 ($\chi^2\sim2$) 
indicates that it is not likely to be a neutron star, and source 9 is situated 
far from the centre of the diffuse emission so is less likely to be the 
associated neutron star. \cha\ sources 10, 11, 12, 13, 15, or 16 in 
Figure~\ref{cha85.4} can also be considered as neutron star candidates (provided 
the optical counterparts listed in Table~\ref{pscat85.4} are not the true counterparts.

\subsection{G85.9$-$0.6}\label{disc85.9}
\begin{figure}
\scalebox{1.0}{\rotatebox{0}{\plotone{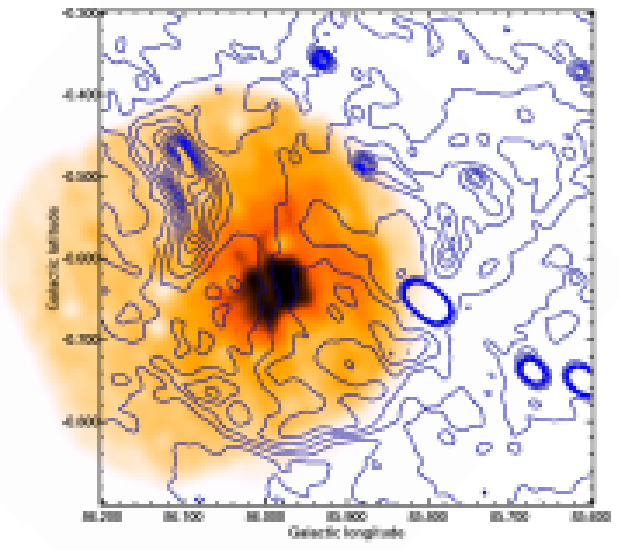}}}
\caption{0.5--2.5 keV \xmm\ X-ray mosaic image of G85.9$-$0.6 with radio
  contours 
  overlaid in blue. The X-ray image has had the point sources
  removed and is Gaussian smoothed with a radius
  of 1$^\prime$ to match the radio resolution, in order to compare the radio
  and diffuse X-ray emission regions. The radio contours were chosen
  to approximate the 
  appearance of the top image in Figure~4 in \cite{kot01}.
\label{85.9imagens}}
\end{figure}

Figure~\ref{85.9imagens} shows X-ray and radio images of
G85.9$-$0.6, produced in a similar way to Figure~\ref{85.4imagens}. The
diffuse X-ray emission appears to contain less structure than
G85.4+0.7, and, as for G85.4+0.7, a spectrum extracted from the central blob, 
the position and size of which are shown in the top panel of Figure~\ref{85.9reg},
indicates that a pulsar wind nebula origin for the emission can be
ruled out. 

Because Figure~\ref{85.4imagens} shows no limb brightening and the
emission is mostly from the central blob, a similar argument to that
used for G85.4+0.7 can be employed here to suggest an ejecta
interpretation for the X-ray emission. Assuming that the free
electrons are evenly distributed in the
extraction area the resulting age is 30 or 22 kyr for the two
backgrounds. Using the average distance between the centre of the
X-ray emission and the shell, $\sim 0.2^\circ$, the radius of the
shell is 15.3 pc  The average expansion velocity would then be 500 or
680 km/s. The emission is concentrated in the inner $6^\prime$, which
indicates an electron density of 0.20 cm$^{-3}$ and an ejecta mass of
1.0 $M_{\sun}$ which agrees with the predicted mass for a type Ia
explosion, 1.4 $M_{\sun}$. This would result in an age of 10600 or
7800 years and average expansion velocity of 1350 or 1850 km/s. 

For a Type Ia supernova, the explosion energy is $\sim10^{51}$ erg and
the ejecta mass is 1.4 $M_{\sun}$. Unlike
for G85.4+0.7, the density in the interarm region is closer to 0.1
cm$^{-3}$, resulting in a swept up mass of nearly 50 $M_{\sun}$. This
indicates that the SNR is in the Sedov expansion phase, which is
described by $R=14 (E_0/n_0)^{1/5}t^{2/5}$, where $R$ is the radius in
pc, $E_0$ is the explosion energy in units of $10^{51}$ erg, $n_0$ is
the ambient density in cm$^{-3}$, and $t$ is the age in units of
$10^4$ years. For an explosion energy of 10$^{51}$ erg, a radius of
15.3 pc, and an age of 10600 or 7800 years, the resulting ambient
density is 0.84 or 0.55 cm$^{-3}$ which are consistent with the
density of the interarm region, but the swept up mass would be 360 or
240 $M_{\sun}$, from which it should be possible to measure thermal
X-ray emission, from the part of the shell that is included in the
X-ray pointing. The current expansion velocity would be
$dR/dt=0.4(R/t)$, which is 570 or 760 km/s.

The interpretation of G85.9$-$0.6 as having been produced by a Type Ia
supernova is reinforced by the Fe abundance, which is well above
solar. As with G85.4+0.7, the ejecta interpretation is called into
question by the remaining abundances, which should be above solar, but
are instead below solar. This could again be a result of poor quality
spectra. 

Given that the radio results described in \cite{kot01}, as well as
the distance presented here, indicated that
G85.9$-$0.6 was most likely produced by a Type Ia supernova, it was not
expected to find a neutron star associated with this SNR. The above
solar Fe abundance for this SNR is consistent with a Type Ia explosion.
An identification of sources 6 and 7 in Figure~\ref{85.9reg} has not
yet been made. They are bright X-ray emitting objects with no known
optical or radio counterpart, making them good neutron star or
radio-quiet AGN candidates, though if one of them were a neutron star,
it would be unlikely that it is associated with the G85.9$-$0.6 SNR
because they are both situated far outside the radio shell of the
remnant, and furthermore, the value of $N_{\rm H}$ for source 6 does not match that
of the SNR itself, and the fit quality of source 7 ($\chi^2\sim2$) indicates that an absorbed 
power law is not a good fit. Their identification with possible 2MASS
counterparts (see Table~\ref{pscat85.9}) makes them more likely radio-quiet AGN than neutron
stars, provided the 2MASS objects are the true counterparts. A future
detailed deep X-ray or multiwavelength study of these objects should be
undertaken to identify and further study them, even though they are
probably not associated with the G85.9$-$0.6 SNR. Source 4
is on the edge of the radio shell of the SNR, but its identification
as a neutron star is questionable because of its photon index and
hardness ratio, as well as the fact that it is not expected that there
is a neutron star associated with this remnant.

\subsection{Mixed Morphology Interpretation}
The centrally filled morphology of both SNRs and the thermal nature of
their X-ray emission confined within the radio shells
suggest that they belong to the class of mixed-morphology SNRs
(also known as thermal composites; Rho \& Petre 1997).
The origin of the thermal X-ray emission interior to the radio
shells in these SNRs has been attributed to several mechanisms which
include: a) cloudlet evaporation in the SNR interior (White \& Long 
1991),
b) thermal conduction smoothing out the temperature gradient across the
SNR and enhancing the central density (Cox et al. 1999), c) a 
radiatively
cooled rim with a hot interior (e.g. Harrus et al. 1997), and d)
possible interaction with a nearby cloud (e.g. Safi-Harb et al. 2005). 
While
modeling these SNRs in the light of the above mentioned models is beyond
the scope of this paper and has to await better quality data, we can 
rule
out the cloudlet evaporation model based on the low ambient densities
inferred from our spectral fits (see \S7.1 and Table~1).  Except for Fe 
and possibly O,
the abundances inferred from our spectral fits are for the most 
elements consistent with
or below solar values, as observed in most mixed-morphology SNRs. 
However, enhanced
metal abundances have been observed in younger SNRs, e.g. 3C~397,
estimated to be $\sim$5.3 kyr--old and proposed to be evolving into the
mixed-morphology phase (Safi-Harb et al. 2005).  The ages inferred
for G85.4+0.7 ($\sim$8--10 kyr; see Table~1 and \S\ref{disc85.4}) and G85.9$-$0.6 ($\sim$8--11 kyrs; 
see Table~1 and \S\ref{disc85.9})
suggest a later evolutionary phase where only shock-heated ejecta from 
Fe
(for G85.9$-$0.6) and possibly Oxygen are still observed.

\acknowledgments

{\it XMM-Newton} is an ESA science mission with instruments and
contributions directly funded by ESA Member States and NASA. This 
research
is supported by the Natural Sciences and Engineering Research Council of
Canada (NSERC), and partly by NASA grant NNG05GL15G. This research has
made use of NASA's Astrophysics Data System, and
the Canadian Galactic Plane Survey, a Canadian project with
international partners supported by NSERC. We thank an anonymous referee for useful comments.

\clearpage
\begin{deluxetable}{lcccc}
\tabletypesize{\scriptsize} 
\tablecaption{Fits of G85.4+0.7 and G85.9$-$0.6 \xmm\ PN and MOS diffuse region spectra to an
  absorbed VPSHOCK and derived quantities. Errors are 2$\sigma$
  uncertainties. Abundances of He, C, and N are frozen to solar.
The Ni abundance is tied to the Fe
  abundance in the fits. \label{dfit}}
\tablewidth{0pt} 
\tablehead{
\colhead{Parameter} & \multicolumn{2}{c}{G85.4+0.7} &
\multicolumn{2}{c}{G85.9$-$0.6}}
\startdata 
MOS Background used:&XMM-ESAS&Region in Figure \ref{85.4reg}&XMM-ESAS&Region in Figure \ref{85.9reg}\\   
\tableline
&&&&\\
$N_{\rm H}$ ($10^{22}$cm$^{-2}$) &$0.86 ^{+0.04}_{-0.05}$& $0.87\pm 0.06$&$0.68^{+0.03}_{-0.04}$&$0.70 \pm 0.03$\\
$kT$ (keV) &$1.1^{+0.8}_{-0.3}$& $1.0^{+0.4}_{-0.2}$&$1.3^{+0.5}_{-0.2}$&$1.6^{+0.5}_{-0.3}$\\
$\tau$ ($10^{10}\rm{cm}^{-3}$s) & $6.4^{+5.8}_{-2.8}$&$8.0^{+8.4}_{-3.4}$&$6.8^{+1.9}_{-0.9}$&$5.1^{+1.5}_{-1.0}$\\
Norm\tablenotemark{a} &$(2.4^{+0.9}_{-0.8})\times10^{-3}$&$(2.3^{+0.8}_{-0.6})\times10^{-3}$ &$(1.1^{+0.1}_{-0.2})\times10^{-3}$&$(1.1\pm0.2)\times10^{-3}$\\
O\tablenotemark{b} & $1.2\pm 0.5$&$1.2^{+2.7}_{-0.6}$ &$1.5^{+0.5}_{-0.4}$ & $1.7^{+0.6}_{-0.4}$\\
Ne & $0.4^{+0.1}_{-0.3}$ &$0.5^{+0.2}_{-0.4}$ & $0.0^{+0.2}_{-0.0}$& $0.0^{+0.2}_{-0.0}$\\
Mg & $0.3\pm0.1$ &$0.4^{+0.1}_{-0.2}$ & $0.6^{+0.2}_{-0.1}$& $0.6^{+0.1}_{-0.2}$\\
Si &$0.1\pm0.1$&$0.2\pm0.2$ &$1.1 \pm0.2$& $0.8\pm0.2$\\
S &$0.4^{+0.2}_{-0.4}$&$0.8^{+1.7}_{-0.8}$&$0.8^{+0.9}_{-0.8}$ &  $0.4^{+0.8}_{-0.4}$\\
Fe & $1.1^{+0.3}_{-0.2}$&$1.1^{+0.5}_{-0.3}$ & $2.6^{+0.5}_{-0.2}$&$2.6\pm0.3$\\
$\chi^2_{\nu}$ ($\nu$)&1.41 (209)&1.38 (211) &1.50 (460)& 1.78 (465)\\
\tableline
&&&&\\
Distance (kpc) & \multicolumn{2}{c}{$3.5\pm 1.0$}&\multicolumn{2}{c}{$4.8\pm1.6$}\\
Radius (pc)\tablenotemark{c}&\multicolumn{2}{c}{($6.4 \pm
  1.8$)$D_{3.5}$}&\multicolumn{2}{c}{($7.5\pm 2.5$)$D_{4.8}$}\\
Temperature (MK) & $13^{+9}_{-3}$&$12^{+5}_{-2}$&$15^{+6}_{-2}$&$19^{+6}_{-3}$\\
0.5--2.5 keV \\
absorbed flux\tablenotemark{d} &$ 1.1\times10^{-12}$ &$ 9.6\times10^{-13}$&$1.5 \times10^{-12}$&$1.6 \times10^{-12}$\\
0.5--2.5 keV \\
unabsorbed flux\tablenotemark{d}&$ 2.7\times10^{-11}$ &$1.8 \times 10^{-11}$ &$1.2 \times10^{-11}$&$1.4 \times10^{-11}$\\
Luminosity \\
(0.5--2.5~keV) (erg/s)\tablenotemark{e} & $(3.1^{+1.3}_{-1.5})\times10^{33}D_{3.5}^2$&$(2.0^{+1.6}_{-1.0})\times10^{33}D_{3.5}^2$ &$(2.4^{+1.2}_{-1.1})\times 10^{34}D_{4.8}^2$& $(2.7^{+1.2}_{-1.3})\times 10^{34}D_{4.8}^2$\\
$n_e$ (cm$^{-3}$)&$(0.11\pm0.03)(fD_{3.5})^{-1/2}$&$(0.11^{+0.05}_{-0.04})(fD_{3.5})^{-1/2}$ &$(0.07^{+0.03}_{-0.02})(fD_{4.8})^{-1/2}$&$(0.07\pm0.03)(fD_{4.8})^{-1/2}$\\
$t$\tablenotemark{f} (kyr) & $(18^{+17}_{-9})(fD_{3.5})^{1/2}$&$(23^{+26}_{-14})(fD_{3.5})^{1/2}$ &$(30^{+12}_{-13})(fD_{4.8})^{1/2}$&$(22^{+11}_{-10})(fD_{4.8})^{1/2}$\\
Mass of X-ray \\
emitting gas ($M_{\sun}$)&$(2.8\pm1.5)f^{1/2}D_{3.5}^{5/2}$&$(2.8^{+1.8}_{-1.6})f^{1/2}D_{3.5}^{5/2}$&$(2.7^{+1.8}_{-1.6})f^{1/2}D_{4.8}^{5/2}$&$(2.7\pm1.8)f^{1/2}D_{4.8}^{5/2}$\\
\enddata 
\tablenotetext{a}{($10^{-14}/4\pi D^2$)$\int n_en_{\rm{H}}dV  \rm{cm}^{-5}$}
\tablenotetext{b}{Elemental abundances are relative to solar abundance.}
\tablenotetext{c}{From X-ray extraction radius in
  Figures~\ref{85.4reg} and \ref{85.9reg}. $D_{3.5}$ and $D_{4.8}$ are distances in terms of 3.5
  and 4.8 kpc, respectively.}
\tablenotetext{d}{Flux in ${\rm erg}\ {\rm cm}^{-2}{\rm s}^{-1}$}
\tablenotetext{e}{$f$ is the volume filling factor.}
\tablenotetext{f}{Shock age}

\end{deluxetable}

\clearpage
\begin{deluxetable}{lcccclcc}
\tabletypesize{\scriptsize} 
\tablecaption{Catalogue of point sources surrounding SNR G85.4+0.7. \cha\ positions are denoted by a CXO prefix\label{pscat85.4}}
\tablewidth{0pt} 
\tablehead{
\colhead{\#} & IAU Name & $\alpha_{\rm J2000.0}$ & $\delta_{\rm
  J2000.0}$ & Pos. Err. &Counterpart\tablenotemark{a} & Offset&$\displaystyle\frac{\rm Optical}{\rm X-ray}$
\\
&&&&(arcsec)&&(arcsec)&Flux Ratio}
\startdata 
1 &XMMU J205101.6+452218& 20 51 01.593 & +45 22 18.47 & 11.9 &1350-13030003 & 1.3&$6.6\times 10^{12}$\\
 & CXO J205101.4+452219 & 20 51 01.392 & +45 22 18.60 & 0.1 & & & \\
 2 &XMMU J205056.8+452322& 20 50 56.750 & +45 23 22.18 & 9.5 & 1350-13027640 & 2.2 &$1.8\times 10^{12}$\\
 & CXO J205056.6+452320 & 20 50 56.603 & +45 23 20.09 & 0.6 & & & \\
 3 &XMMU J205043.2+452213& 20 50 43.212 & +45 22 12.96 & 11.2 & 1350-13020622 &  2.3&$6.1\times 10^{12}$\\
 & CXO J205043.1+452215 & 20 50 43.097 & +45 22 15.49 & 0.8 & & & \\
 4 &XMMU J205034.8+451831& 20 50 34.811 & +45 18 31.47 & 10.1 & 1350-13016264 & 2.9&$6.8\times 10^{14}$\\
 & CXO J205034.9+451836 & 20 50 34.945 & +45 18 35.83 & 1.2 & & & \\
 5 &XMMU J205024.5+452343& 20 50 24.523 & +45 23 42.95 & 11.9 & 1350-13011037 & 2.7&$2.7\times 10^{12}$\\
 6 &XMMU J205120.9+451859& 20 51 20.925 & +45 18 58.73 & 12.4 & J205120.79+451900.1(S)\tablenotemark{b} & 2.4&$1.2\times 10^{15}$\\
 & CXO J205120.8+451901 & 20 51 20.802 & +45 19 01.11 & 0.4 & & & \\
\tablenotemark{c}&&&&& 1350-13039935&4.1&$2.6\times 10^{15}$\\
\tablenotemark{d}&&&&& &5.1&\\
7 &XMMU J205127.2+452025& 20 51 27.152 & +45 20 24.96 & 11.6 & 1350-13042959 & 1.7&$4.6\times 10^{15}$\\
 & CXO J205127.2+452025 & 20 51 27.158 & +45 20 25.01 & 0.6 & & & \\
 8 &XMMU J205000.7+452044& 20 50 00.657 & +45 20 43.94 & 12.2 & J205000.64+452046.2(S) & 2.0&$1.4\times 10^{15}$ \\
&&&&&1350-12998831&3.7&$1.4\times 10^{15}$\\
 9 &XMMU J204959.8+452349& 20 49 59.800 & +45 23 49.00 & 14.5 & 1350-12998401& 1.7&$4.3\times 10^{12}$\\
&&&&&1353-0399326(B)&2.1&$6.8\times 10^{11}$\\
10 & CXO J205109.0+452440 & 20 51 09.022 & +45 24 39.85 & 1.4 &1350-13033336 &11.2 &$ 2.3\times 10^{13}$    \\
11 & CXO J205056.6+451744 & 20 50 56.588 & +45 17 44.11 & 1.0 &1350-130274980& 6.3&$ 1.2\times 10^{12}$    \\
12 & CXO J205111.9+452136 & 20 51 11.876 & +45 21 36.26 & 1.0 &1350-13035376 &1.7 &$ 2.1\times 10^{14}$    \\
13 & CXO J205036.9+451945 & 20 50 36.913 & +45 19 45.00 & 1.1 &1350-13017472 &7.8 &$ 1.4\times 10^{12}$    \\
14 & CXO J205138.4+452333 & 20 51 38.370 & +45 23 33.43 & 1.2 &1350-13049047 &6.0 &$ 3.3\times 10^{12}$    \\
15 & CXO J205115.5+452207 & 20 51 15.499 & +45 22 07.42 & 1.8 &1353-0400904(B) &6.2 &$ 8.4\times 10^{11}$    \\
16 & CXO J205103.6+451730 & 20 51 03.641 & +45 17 30.00 & 1.7 &1350-13031016 &3.8 &$ 2.5\times 10^{12}$    \\
17 & CXO J205149.3+452341 & 20 51 49.264 & +45 23 40.59 & 1.7 &1350-13054386 &2.4 &$ 1.0\times 10^{12}$    \\
18 & CXO J205140.0+452107 & 20 51 40.002 & +45 21 06.64 & 2.0 &1350-13049401 &5.6 &$ 4.9\times 10^{11}$    \\
19 & CXO J205143.3+452037 & 20 51 43.338 & +45 20 37.04 & 1.7 &1350-13051265 &3.2 &$ 3.7\times 10^{12}$    \\
20 & CXO J205135.2+451830 & 20 51 35.164 & +45 18 30.27 & 1.7 &1350-13047169 &1.5 &$ 6.4\times 10^{15}$    \\
21 & CXO J205126.4+451856 & 20 51 26.420 & +45 18 56.06 & 1.6 &1353-0401091 &7.2 &$ 1.8\times 10^{12}$    \\
\enddata  
\tablenotetext{a}{All designations are from the USNO-A2.0 catalogue \citep{mon98} unless otherwise noted.}
\tablenotetext{b}{(S) denotes SKY2000 designation \citep{my02}}
\tablenotetext{c}{This row refers to the \xmm\ position}
\tablenotetext{d}{This row refers to the \cha\ position}

\end{deluxetable}

\clearpage
\begin{deluxetable}{lcccclcc}
\tabletypesize{\scriptsize} 
\tablecaption{Catalogue of point sources surrounding SNR G85.9$-$0.6. \cha\ positions are denoted by a CXO prefix\label{pscat85.9}}
\tablewidth{0pt} 
\tablehead{
\colhead{\#} & IAU Name & $\alpha_{\rm J2000.0}$ & $\delta_{\rm
  J2000.0}$ & Pos. Err. &Counterpart\tablenotemark{a} & Offset&$\displaystyle\frac{\rm Optical}{\rm X-ray}$
\\
&&&&(arcsec)&&(arcsec)&Flux Ratio}
\startdata 
1 &XMMU J205911.4+444729&20 59 11.398& +44 47 29.27& 11.3& 1347-0402736(B)\tablenotemark{b}&  5.5&$1.1\times 10^{19}$    \\
2 &XMMU J205950.5+445954&20 59 50.504& +44 59 53.58& 10.3& HD 200102(I)\tablenotemark{c}&6.0&$3.7\times 10^{15}$\\
3 &XMMU J205944.3+450741&20 59 44.261& +45 07 40.75& 11.7& 1350-13269810        & 3.7 &$1.1\times 10^{14}$\\
4 &XMMU J205857.0+450348&20 58 56.993& +45 03 48.25& 11.0& 1350-0403688(B)&  3.7&$1.0\times 10^{13}$\\
5 &XMMU J210003.3+450342&21 00 03.280& +45 03 41.76& 10.4& 1350-13276811&  4.3&$2.5\times 10^{12}$\\
6 &XMMU J210023.1+445435&21 00 23.149& +44 54 34.98& 11.1& 21002305+4454364(M)\tablenotemark{d}&  1.4&\nodata\\
7 &XMMU J210000.1+445631&21 00 00.057& +44 56 30.94&  9.3& 21000013+4456364(M) &  4.9&\nodata   \\
8 &XMMU J205917.0+445809&20 59 17.009& +44 58 08.71& 10.5& 1349-0404180(B)&  1.4&$4.2\times 10^{12}$\\
 & CXO J205917.3+445821 & 20 59 17.308 & +44 58 20.90 & 0.5 & & 12.2& \\
\tablenotemark{e}&&&&& 1349-0404191(B)&11.7&$2.7\times 10^{14}$\\
\tablenotemark{f}&&&&& &1.1&\\
\tablenotemark{e}&&&&& 1275-14437760&12.3&$5.5\times 10^{13}$\\
\tablenotemark{f}&&&&& &0.7&\\
9 & CXO J205915.7+445858 & 20 59 15.715 & +44 58 57.64 & 1.1 &1275-14437093 &9.6 & $6.2\times 10^{14}$\\
10 & CXO J205849.8+445725 & 20 58 49.768 & +44 57 25.00 & 0.8 &1275-14424519 &2.7 &$8.2 \times 10^{13}$\\
11 & CXO J205839.9+445132 & 20 58 39.924 & +44 51 31.67 & 1.3 &1275-14419711 &5.4 &$1.8 \times 10^{15}$\\
12 & CXO J205835.3+444640 & 20 58 35.349 & +44 46 40.21 & 1.2 &1347-0402278(B) &1.1 &$8.4\times 10^{11}$\\
13 & CXO J205825.4+444940 & 20 58 25.398 & +44 49 39.63 & 1.7 &1348-0403023(B) & 7.1&$6.1 \times 10^{14}$\\
14 & CXO J205823.9+444621 & 20 58 23.867 & +44 46 21.35 & 1.7 &1347-0402123(B) &3.2 &$1.8 \times 10^{12}$\\
&&&&& 20582367+4446195(M)&1.8&\nodata\\
15 & CXO J205930.8+445729 & 20 59 30.760 & +44 57 29.19 & 1.1 &1349-0404462(B) &4.0 &$2.9 \times 10^{13}$\\
16 & CXO J205922.2+445416 & 20 59 22.191 & +44 54 15.95 & 1.5 &1349-0404292(B) &3.1 &$1.8 \times 10^{13}$\\
&&&&& 20592216+4454176(M)&1.7&\nodata\\
17 & CXO J205921.9+445818 & 20 59 21.928 & +44 58 18.35 & 1.1 & 1275-14439638&8.6 &$2.6 \times 10^{12}$\\
&&&&& 20592206+4458188(M)&1.6&\nodata\\
18 & CXO J205920.0+445321 & 20 59 19.993 & +44 53 21.12 & 1.3 &1275-14439333 &7.0 &$7.0 \times 10^{13}$\\
&&&&& 20592022+4453178(M)&4.1&\nodata\\
19 & CXO J205912.6+445033 & 20 59 12.619 & +44 50 33.10 & 1.1 &20591251+4450356(M) & &\nodata\\
20 & CXO J205904.1+445952 & 20 59 04.133 & +44 59 52.38 & 1.1 &1275-14431663 &7.8 & $3.5\times 10^{13}$\\
21 & CXO J205857.7+450008 & 20 58 57.717 & +45 00 07.51 & 1.2 &\nodata & \nodata&\nodata\\
22 & CXO J205851.5+450050 & 20 58 51.450 & +45 00 50.04 & 1.3 &1350-0403625(B) &3.5 &$2.4\times 10^{12}$\\
23 & CXO J205848.7+445920 & 20 58 48.681 & +44 59 19.84 & 1.2 &1275-14423599 &8.4 &$2.3 \times 10^{13}$\\
24 & CXO J205841.1+445260 & 20 58 41.091 & +44 52 59.50 & 1.1 &\nodata &\nodata & \nodata\\
25 & CXO J205837.4+445510 & 20 58 37.389 & +44 55 09.86 & 1.1 &1275-14418321 &9.1 & $1.0\times 10^{13}$\\
&&&&& 20583749+4455092(M)&1.3&\nodata\\
26 & CXO J205834.6+445330 & 20 58 34.649 & +44 53 30.33 & 1.4 &1348-0403158(B) & & $3.2\times 10^{12}$\\
27 & CXO J205814.7+444606 & 20 58 14.652 & +44 46 05.70 & 2.1 &\nodata &\nodata & \nodata\\
\enddata  
\tablenotetext{a}{All designations are from the USNO-A2.0 catalogue \citep{mon98} unless otherwise noted.}
\tablenotetext{b}{(B) denotes USNO-B1.0 designation \citep{mon03}}
\tablenotetext{c}{(I) denotes designation in \cite{gua92}}
\tablenotetext{d}{(M) denotes 2MASS designation \citep{2mass}}
\tablenotetext{e}{This row refers to the \xmm\ position}
\tablenotetext{f}{This row refers to the \cha\ position}

\end{deluxetable}
\clearpage
\begin{deluxetable}{lcccccccc}
\tabletypesize{\scriptsize} 
\tablecaption{Parameters for fits of \xmm\ PN, MOS and \cha\tablenotemark{a} point source spectra to an absorbed power law. \label{psspectab}}
\tablewidth{0pt} 
\tablehead{
\colhead{\#} & $N_{\rm H}$ ($10^{22}$cm$^{-2}$) & $\Gamma$ & PL
Normalization\tablenotemark{b} & $\chi^2_{\nu}$ ($\nu$)&0.5--2.0 keV &
2.0--10.0 keV&Hardness ratio\tablenotemark{c} &X-ray 
\\
&&&&&Counts&Counts&&S/N}
\startdata 
\multicolumn{3}{l}{G85.4+0.7}&&&&\\
&&&&&&&&\\
1&$0.50^{+0.25}_{-0.18}$&$2.47^{+0.63}_{-0.47}$&$4.75^{+3.5}_{-1.8}$&1.20 (37)&210 & 56&0.27 & 3.8    \\
2& $0.94^{+0.27}_{-0.56}$&$8.7^{+1.3}_{-3.3}$&$11.2^{+11.8}_{-8.2}$&0.93 (14)&122 & 0 & 0.00           & 7.2  \\
3&   $0.70^{+1.13}_{-0.69}$&$4.9\pm3.1$&$4.9^{+74.1}_{-4.2}$&0.75 (12)&  92 & 0& 0.00         & 6.2 \\
4&$0.3^{+0.8}_{-0.2}$&$3.8^{+6.5}_{-1.5}$&$2.8^{+7.2}_{-2.4}$&1.10 (13)& 150 & 50& 0.33          & 4.8      \\
5&$1.1^{+0.1}_{-1.0}$&$9.5^{+0.5}_{-5.4}$&$13.2^{+3.4}_{-12.8}$&2.32 (4)& 83 & 0& 0.00           & 4.5     \\
6&$0.89^{+0.09}_{-0.18}$&$9.5^{+0.5}_{-1.2}$&$16.5^{+6.8}_{-6.5}$&5.58 (24)&  139 & 0& 0.00          & 4.8  \\
7&$1.19^{+0.48}_{-0.99}$&$8.7^{+1.3}_{-5.5}$&$17.3^{+32.7}_{-16.5}$&0.73 (5)& 104 & 0& 0.00                & 9.0  \\
8&$0.93^{+0.19}_{-0.73}$&$9.5^{+0.5}_{-3.4}$&$14^{+9}_{-13}$&1.17 (10)& 96 & 0& 0.00             & 9.2   \\
9&$0.75^{+0.84}_{-0.31}$&$2.6^{+1.4}_{-0.8}$&$6.4^{+16.6}_{-4.2}$&0.939 (17)& 86 & 43&0.33                    & 3.2   \\
\tableline
&&&&&&&&\\
\multicolumn{3}{l}{G85.9$-$0.6}&&&&\\
&&&&&&&\\
1& $1.04^{+0.03}_{-0.06}$& $9.5^{+0.5}_{-0.3}$&$117^{+14}_{-20} $&7.73 (35)& 1230&20&0.016           & 35.4     \\
2& $1.1^{+0.1}_{-0.8}$& $8.5^{+1.3}_{-4.4}$&$53^{+21}_{-47}$ &3.87 (22)& 433&2 &0.005             & 20.9     \\
3&  $0.86^{+0.05}_{-0.12}$&$9.5\pm 0.5$&$32^{+9}_{-11}$&11.12 (18)& 417&0 & 0.00                 & 20.4     \\
4& $0.53^{+0.77}_{-0.28}$& $3.9^{+3.8}_{-1.3}$&$4.4^{+30}_{-2.9}$&2.13 (10)& 177& 0& 0.00   & 13.3     \\
5& $0.85^{+0.21}_{-0.79}$& $8.9^{+1.1}_{-4.3}$& $14^{+19}_{-13}$&1.27 (6)& 117&18&0.15                & 11.6     \\
6& $1.32^{+0.76}_{-0.52}$&$1.56^{+0.62}_{-0.47}$ & $9.9^{+12.5}_{-4.9}$&1.07 (19)&112& 301&2.69           & 20.3     \\
7& $0.71^{+1.06}_{-0.50}$&$1.70^{+1.43}_{-0.77}$ & $2.5^{+9.1}_{-9.2} $& 2.19 (10)&26 & 90&3.46      & 10.8     \\
8& $0.46^{+0.30}_{-0.21}$& $5.3^{+1.8}_{-1.3}$&$5.0 ^{+11.7}_{-2.4}$ & 1.71 (33)&473 &30 & 0.06             & 23.1     \\
14&$$&$1.3_{-1.3}^{+4.0}$&$0.6_{-0.2}^{+8.0}$&
23
26
\enddata  
\tablenotetext{a}{\cha\ spectra are included in fit for sources 1, 2, 3, 4, 6, and 7 of G85.4+0.7 and source 8 of G85.9$-$0.6.}
\tablenotetext{b}{10 $^{-5}$ photons keV$^{-1}$cm$^{-2}$s$^{-1}$ at 1 keV}
\tablenotetext{c}{Hardness ratio is the ratio of hard counts to total (hard+soft) counts}

\end{deluxetable}

\clearpage
 
\clearpage 

\clearpage  

\clearpage

\clearpage  

\clearpage  

\clearpage

\clearpage  

\clearpage  

\clearpage  

\clearpage  

\clearpage  

\clearpage

\clearpage  

\clearpage

\clearpage


\begin{thebibliography}{}
\bibitem[Borkowski et al.(2001)]{bor01} Borkowski, K. J., Lyerly,
    W. J., \& Reynolds, S. P. 2001 \apj, 548, 820
\bibitem[Buccheri et al.(1983)]{buc83} Buccheri, R., et al. 1983,
  A\&A, 128, 245
\bibitem[Corcoran et al.(2000)]{cor00} Corcoran, M. F., Fredericks,
  A. C., Petre, R., Swank, J. H., \& Drake, S. A. 2000, \apj, 545, 420
\bibitem[Cox et al.(1999)]{cox99} Cox, D., Shelton, R. L., Maciejewski, W., Smith, R. K., Plewa, T., Pawl, A., \& Ryczka, M. 1999, \apj, 524, 179
\bibitem[Dame et al.(1987)]{dame87} Dame, T.~M. et al. 1987, \apj,
  322, 706
\bibitem[Dame et al.(2001)]{dame01} Dame, T.~M., Hartmann, D., \&
  {Thaddeus}, P. 2001, \apj, 547, 792
\bibitem[Eisenhauer et al(2005)]{eise05} Eisenhauer, F., et al. 2005,
\apj, 628, 246
\bibitem[Foster \& MacWilliams(2006)]{fmac06} Foster, T. \& MacWilliams,
J. 2006, \apj, 644, 214
\bibitem[Gotthelf(2006)]{gott06} Gotthelf, E. V. 2006, in ``Young Neutron Stars and Their Environments'' (IAU Symposium 218, ASP Conference Proceedings), eds. F. Camilo and B. M. Gaensler. 225
\bibitem[Guarinos(1992)]{gua92}Guarinos J. 1992, Distribution of
  interstellar matter in the galactic disk from
visual extinction data, in ``Astronomy from Large Databases II'',
  Haguenau 14-16 September 1992,
      Ed. A. Heck and F. Murtagh, ESO Conference and Workshop Proceedings No 43,
      ISBN 3-923524-47-1, p. 301
\bibitem[Harrus et al.(1999)]{har99} Harrus, I. M., Hughes, J. P., Singh, K. P., Koyama, \& Asaoka, I. 1999, \apj, 488, 781
\bibitem[Jackson, Safi-Harb, \& Kothes(2006)] {jac06} Jackson, M., Safi-Harb, S., \&
Kothes, R. 2006 Canadian Astronomical Society Meeting, Calgary, Canada, June 1-4
\bibitem[Kothes et al.(2001)]{kot01} Kothes, R., Landecker, T. L.,
    Foster, T., and Leahy, D. A. 2001, A\&A, 376, 641
\bibitem[Leahy, Elsner, \& Weisskopf (1983)]{lea83} Leahy, D. A.,
  Elsner, R. F., \& Weisskopf, M. C. 1983, \apj, 272, 256
\bibitem[Liedahl et al.(1990)]{lie90} Liedahl, D. A., Kahn, S. M.,
  Osterheld, A. L., \& Goldstein, W. H. 1990, \apj, 350, L37
\bibitem[Lumb(2002)]{lum02} Lumb, D. 2002, EPIC Background Files,
  XMM-SOC-CAL-TN-0016, Issue 2.0, http://xmm.vilspa.esa.es/docs/documents/CAL-TN-0016-2-0.ps.gz
\bibitem[Lyne \& Graham-Smith(1998)]{lyn98} Lyne, A. G., \&
  Graham-Smith, F. 1998, Pulsar Astronomy, Cambridge: Cambridge
  University Press.
\bibitem[Mewe et al.(1985)]{mew85} Mewe, R., Gronenschild,
  E. H. B. M., \& van den Oord, G. H. J. 1985, A\&AS, 62, 197
\bibitem[Mewe et al.(2003)]{mew03} Mewe, R., Raasssen, A. J. J.,
  Cassinelli, J. P., van der Hucht, K. A., Miller, N. A., \& G{\"u}del,
  M. 2003 A\&A, 398, 203
  ,\& van den Oord, G. H. J. 1985, A\&AS, 62, 197
\bibitem[Monet et al.(1998)]{mon98} Monet, D., et al. 1998, USNO-A V2.0, A Catalog of Astrometric Standards
\bibitem[Monet et al.(2003)]{mon03} Monet, D. G., et al. 2003, \aj, 125, 948
\bibitem[Morrison \& McCammon (1983)]{morr83} Morrison, R., \& McCammon, D. 1983, \apj, 270, 119
\bibitem[Myers et al.(2002)]{my02}Myers J.R., Sande C.B., Miller
  A.C., Warren Jr. W.H., Tracewell D.A. 2002, SKY2000 Master Catalog, Version 4
\bibitem[Read \& Ponman(2003)]{rea03} Read A.M. \& Ponman T.J., 2003, A\&A, 409, 395
\bibitem[Rho \& Petre(1997)]{rho97} Rho, J., \& Petre, R. 1997, \apj, 484, 828
\bibitem[Roberts (1969)]{robe69} Roberts, W. W. 1969, \apj, 158, 123
\bibitem[Rohlfs \& Wilson(2004)]{roh04} Rohlfs, K. \& Wilson. T. L. 2004, ``Tools of Radio Astronomy'', 4th rev. \& enl. ed., Berlin: Springer, 2004
\bibitem[Safi-Harb(2006)] {saf06} Safi-Harb, S. 2006, American Astronomical Society, 208, \#61.03
\bibitem[Safi-Harb et al.(2005)]{saf05} Safi-Harb, S., Dubner, G., Petre, R., Holt, S. S., \& Durouchoux, P. 
2005,
\apj, 618, 321
\bibitem[Skrutskie et al.(2006)]{2mass} Skrutskie, M. F., et al.
    2006, AJ, 131, 1163
\bibitem[Snowden \& Kuntz(2006)]{snow06} Snowden, S. L. \& Kuntz,
  K. D., 2006. Cookbook for Analysis Procedures for \xmm\ EPIC MOS
  Observations of Extend Objects and the Diffuse Background, Version 1.0.1.
\bibitem[Str\"{u}der et al.(2001)]{str01} Str\"{u}der,  L., et al. 2001, A\&A, 365, L18
\bibitem[Taylor et al.(2003)]{tay03} Taylor, A.~R. et al. 2003, \aj, 125, 3145.
\bibitem[Townsley et al.(2000)]{tow00} Townsley, L. K., Broos, P. S., Garmire, G. P., Nousek, J. A., 2000, \apj, 534, L139
\bibitem[Turner et al.(2001)]{tur01} Turner,  M. J. L., et al. 2001, A\&A, 365, L27
\bibitem[White \& Long(1991)]{whi91} White, R. L.  \& Long, K. S. 1991, \apj, 373, 543
\bibitem[Wielen (1979)]{wiel79} Wielen, R. 1979, in IAU Symp. 84, The
Large-Scale Characteristics of the Galaxy, ed. W. B. Burton (Dordrecht:
Reidel), 133
\bibitem[Woods \& Thompson(2006)]{wood04} Woods, P. \& Thompson, C. 2006 in ``Compact Stellar X-ray Sources'', eds. W. Lewin \& M. van der Klis (Cambridge: Cambridge University Press), 547 
\end{thebibliography}
\end{document}